\documentstyle[preprint,aps,eqsecnum,amssymb,amsfonts,verbatim]{revtex}
\tightenlines

\begin{document}

\title{\bf Rigged Hilbert Space Approach to the Schr\"odinger Equation}

\author{R.~de la Madrid}

\address{Institute for Scientific Interchange (ISI),
Villa Gualino, Viale Settimio Severo 65, I-10133, Torino, Italy \\
and \\
Departamento de F\'\i sica Te\'orica, Facultad de Ciencias, 47011 Valladolid, 
Spain \\
E-mail: \texttt{rafa@isiosf.isi.it}}

\date{October 1, 2001}

\maketitle

\begin{abstract}
\noindent It is shown that the natural framework for the solutions of any 
Schr\"odinger equation whose spectrum has a continuous part is the Rigged 
Hilbert Space rather than just the Hilbert space.  The difficulties of using
only the 
Hilbert space to handle unbounded Schr\"odinger Hamiltonians whose spectrum 
has a continuous part are disclosed.  Those difficulties are overcome by using
an appropriate Rigged Hilbert Space (RHS).  The RHS is able to associate an 
eigenket to each energy in the spectrum of the Hamiltonian, regardless of
whether the energy belongs to the discrete or to the continuous part of the
spectrum.  The collection of eigenkets corresponding to both discrete
and continuous spectra forms a basis system that can be used to expand any 
physical wave function.  Thus the RHS treats discrete energies (discrete
spectrum) and scattering energies (continuous spectrum) on the same 
footing. 
\end{abstract}


%PACS numbers: 03.65.-w, 02.30.-f

\def\llra{\relbar\joinrel\longrightarrow}              %THIS IS LONG
\def\mapright#1{\smash{\mathop{\llra}\limits_{#1}}}    %ARROW ON LINE
\def\mapup#1{\smash{\mathop{\llra}\limits^{#1}}}     %CAN PUT SOMETHING OVER IT
\def\mapupdown#1#2{\smash{\mathop{\llra}\limits^{#1}_{#2}}} %over&under it%

\newpage

\def\thesection{\arabic{section}}
\section{Introduction}
\def\thesection{\arabic{section}}
\setcounter{equation}{0}
\label{sec:introduction}

Several authors have realized that the Hilbert space is not sufficient for 
the purposes of Quantum Mechanics, and that an 
extension of the Hilbert space to the Rigged Hilbert Space (RHS) is 
needed.  The RHS was introduced in physics for the first time in the 1960s 
independently by J.~P.~Antoine~\cite{ANTOINE}, A.~Bohm~\cite{B60}, and 
J.~E.~Roberts~\cite{ROBERTS}.  These authors realized that the 
RHS provides a rigorous mathematical rephrasing of Dirac's formalism.  In 
essence, the Nuclear Spectral Theorem~\cite{GELFAND} (also known as 
Gelfand-Maurin Theorem) restates Dirac basis vector expansion along with the 
Dirac bras and kets within a mathematical theory.  Later on, other authors such
as A.~Galindo and P.~Pascual~\cite{GALINDO}, and O.~Melsheimer~\cite{MELSH} 
came to the same conclusion.  

Earlier attempts to go beyond the Hilbert space framework are reported in
Refs.~\cite{MEJLBO,KRISTENSEN}.  If two operators of the algebra of 
observables satisfy Heisenberg's commutation relation, at least 
one of them cannot be continuous (bounded) with respect to the Hilbert space 
topology.  In Refs.~\cite{MEJLBO,KRISTENSEN}, it is shown that there are
subdomains of the Hilbert space that can be endowed with 
topologies that make those operators continuous.

Since the early 1990s, the RHS has become a standard tool in many areas of
theoretical physics, especially in those that deal with continuous and
resonance spectra.  For instance, the RHS has been used to treat the 
Lippmann-Schwinger equation~\cite{DIS,BKR}, the Gamow vectors 
(cf.~\cite{DIS,BOLLINI96,BOHM97,ANTOINE98,KUKULIN} and references 
therein), and certain generalized spectral decompositions of chaotic 
maps~\cite{AT93,SUCHANECKI}. 

The dynamical equation of Quantum Mechanics is the Schr\"odinger 
equation.  Thus any attempt to show that the RHS contains the mathematical 
methods needed by Quantum Mechanics should show that the natural framework for 
the solutions of the Schr\"odinger equation is the RHS.  The objective of this
paper is to show that the solutions of the Schr\"odinger equation fall in a 
RHS.  To illustrate this, we shall use the example of the square well-barrier 
potential Hamiltonian.  We recall that none of 
Refs.~\cite{BKR,BOLLINI96,BOHM97,ANTOINE98,KUKULIN,AT93,SUCHANECKI} 
took the Schr\"odinger equation as the dynamical equation.

A RHS is a triplet of spaces
\begin{equation}
      \mathbf \Phi \subset {\cal H} \subset \mathbf \Phi ^{\times}  \, ,
     \label{IRHS}
\end{equation}
where $\cal H$ is a Hilbert space, $\mathbf \Phi$ is a dense subspace of the
Hilbert space, and $\mathbf \Phi ^{\times}$ is the dual space of
$\mathbf \Phi$, i.e., $\mathbf \Phi ^{\times}$ is the set of antilinear
functionals over the space $\mathbf \Phi$ .  The space $\mathbf \Phi$ has a 
topology that is finer than the topology inherited from $\cal H$.  The domain 
${\cal D}(H)$ of the Hamiltonian lies between $\mathbf \Phi$ and $\cal H$,
\begin{equation}
       {\mathbf \Phi} \subset {\cal D}(H) \subset {\cal H} \, .
\end{equation}

If a Schr\"odinger Hamiltonian is defined on the whole Hilbert space and has
only discrete spectrum, then the Hilbert space is sufficient for the purposes 
of Quantum Mechanics.  However, if the Hamiltonian is not defined on the
whole Hilbert space (e.g., it is unbounded) and its spectrum has a 
continuous part, then the mathematical methods of the Hilbert space are not 
sufficient, and an extension of those methods is needed.  The RHS arises as
the natural extension when dealing with unbounded operators that have 
continuous spectrum.

The reason why the RHS is the extension of the Hilbert space that we need
is twofold.  On the one hand, one of the key assumptions of Quantum Mechanics 
is that the quantity
\begin{equation}
       (\varphi ,H\varphi ) 
        \label{IexintrodispP}
\end{equation}
accounts for the expectation value of the measurement of the observable
$H$ in the state $\varphi$, and that
\begin{equation}
      \Delta _{\varphi}H=
      \sqrt{ (\varphi ,H^2\varphi )-(\varphi ,H\varphi )^2}
      \label{IintrodispA}
\end{equation}
accounts for the uncertainty of the measurement of the observable $H$ in the
state $\varphi$.  The expectation value (\ref{IexintrodispP}) cannot be 
computed for every element of the Hilbert space $\cal H$, but only for those 
$\varphi \in {\mathcal H}$ that also belong to ${\cal D}(H)$.  Similarly, the 
uncertainty (\ref{IintrodispA}) cannot be computed for every element of 
$\cal H$ either.  If we take as 
physical states those square integrable functions for which physical 
quantities such as the expectation value (\ref{IexintrodispP}) and the 
uncertainty (\ref{IintrodispA}) can be computed, then not every square 
integrable function (i.e., every element of $\cal H$) can represent a 
physical state.  As we shall see, the natural space of physical wave functions
is the space $\mathbf \Phi$ in (\ref{IRHS}), because all physical 
quantities such as expectation values and uncertainties can be computed for 
its elements.  

On the other hand, in Quantum Mechanics it is assumed that for each energy 
in the spectrum of the Hamiltonian $H$, there corresponds a ket that is an 
eigenvector of $H$.  If we denote the ket corresponding to an energy $E_n$ in 
the discrete spectrum by $|E_n)$ and denote the ket 
corresponding to an energy $E$ in the continuous spectrum
by $|E\rangle$, then we should have
\begin{mathletters}
      \label{Ieigenkets}
\begin{eqnarray}
      \def\theequation{\thesection.\arabic{equation}}
      &H|E_n)=E_n|E_n) \, , \quad &  
         \label{Ibeigenkets}  \\
      &H|E\rangle=E|E\rangle \, . \quad & 
        \label{Iceigenkets}  
\end{eqnarray} 
\end{mathletters}
These eigenkets are normalized according to the following rule:
\begin{equation}
       (E_n|E_m)=\delta _{nm} \, , \quad    \langle E |E ^{\prime} \rangle =
           \delta (E -E ^{\prime}) \, , \quad \langle E|E_n)=0 \, .
         \label{Ideltanorintro}
\end{equation}
It is also assumed that those kets form a complete basis that can be used to 
expand any wave function $\varphi$,
\begin{equation}
       \varphi = \sum_{n}|E_n)(E_n|\varphi ) +\int dE \,
             |E\rangle \langle E|\varphi \rangle \, .
       \label{IDiracbsis}
\end{equation}
The solutions of Eq.~(\ref{Ibeigenkets}) are square normalizable, i.e., 
they lie in the Hilbert space.  However, the solutions of 
Eq.~(\ref{Iceigenkets}) are not square normalizable, i.e., they lie outside
the Hilbert space.  Therefore, the Hilbert space is not large enough to 
contain the non-normalizable eigenkets that are associated to the energies in 
the continuous spectrum---a larger space than the Hilbert space is 
needed.  This larger space, which contains the eigenkets $|E\rangle$, is
the space $\mathbf \Phi ^{\times}$ in (\ref{IRHS}).  The action of the 
Hamiltonian, which is in principle defined only on the elements of 
$\mathbf \Phi$ (or on the elements of ${\cal D}(H)$), can be extended to the 
elements $|F\rangle$ of $\mathbf \Phi ^{\times}$ by defining the following 
conjugate operator $H^{\times}$:  
\begin{equation}
      \langle \varphi |H^{\times}|F\rangle :=
        \langle H^{\dagger}\varphi |F\rangle  \, ,
      \quad \forall \varphi \in {\mathbf \Phi} \, , \ |F\rangle \in
       \mathbf \Phi ^{\times} \, .
       \label{idef}
\end{equation}
The operator $H^{\times}$ is a uniquely defined extension of $H$.  Using
definition (\ref{idef}), the RHS formalism restates Eq.~(\ref{Iceigenkets}) as 
\begin{equation}
      \langle \varphi |H^{\times}|E\rangle =
      \langle H^{\dagger}\varphi |E\rangle =E\langle \varphi |E\rangle 
        \, , \quad 
      \forall \varphi \in \mathbf \Phi \, .
      \label{igeing}
\end{equation}
When the arbitrary $\varphi \in \mathbf \Phi$ is omitted in this equation,
we recover the formal equation (\ref{Iceigenkets}).

In this way, the Gelfand triplet (\ref{IRHS}) arises in a natural way.  The 
Hilbert space $\mathcal H$ appears because the wave functions must be square 
normalizable.  The subspace $\mathbf \Phi$ is the set
of physical wave functions, i.e., the set of square integrable functions for 
which any expectation
value and any uncertainty can be computed.  The dual space
$\mathbf \Phi ^{\times}$ contains the eigenkets associated to energies in the 
continuous part of the spectrum of $H$.  These eigenkets are defined as 
functionals over the space $\mathbf \Phi$, and they can be used to expand any 
$\varphi \in \mathbf \Phi$ as in Eq.~(\ref{IDiracbsis}).  The eigenequation 
(\ref{Iceigenkets}) holds within the RHS in the sense of 
Eq.~(\ref{igeing}).  Thus the Rigged Hilbert Space is the extension of the 
Hilbert space that is needed when the solutions of the Schr\"odinger equation 
are not normalizable.  These non-normalizable solutions are treated as 
distributions in the dual space $\mathbf \Phi ^{\times}$.

Unlike the Hilbert space, the RHS treats the discrete and continuous spectra 
on the same footing: for each energy of the spectrum of the Hamiltonian 
(either in the discrete or in the continuous part), there is an eigenket of 
the Hamiltonian (given by Eq.~(\ref{Ibeigenkets}) in the discrete case, and by
Eq.~(\ref{Iceigenkets}) in the continuous case) that belongs to a basis that 
expands any physical wave function as in Eq.~(\ref{IDiracbsis}).

The main shortcoming of the RHS theory is that it does
not provide a prescription to construct the space $\mathbf \Phi$ and the 
eigenvectors $|E\rangle$.  The general statement of the Nuclear Spectral 
Theorem~\cite{GELFAND} just assures the existence of the generalized 
eigenvectors $|E\rangle$.  In that theorem, the space $\mathbf \Phi$ is 
assumed to be given beforehand.  In this paper, the RHS of the square
well-barrier will be constructed explicitly along with the eigenkets
$|E\rangle$ by applying the method proposed in Ref.~\cite{DIS}.  This method
has been successfully applied to a Hamiltonian with purely continuous 
spectrum: the square {\it barrier} Hamiltonian~\cite{FP}.  Here we apply it 
to a Hamiltonian with both discrete and continuous spectrum: the square 
{\it well-barrier} Hamiltonian.  

The steps of our method can be summarized as follows:  

\vskip0.2cm

(i) Construction of the self-adjoint Hamiltonian from the formal 
Schr\"odinger differential operator.

\vskip0.2cm

(ii) Construction of the Green function and therewith of the resolvent.
 
\vskip0.2cm

(iii) Computation of the spectrum of the Hamiltonian.

\vskip0.2cm

(iv) Construction of the direct integral decomposition.
      
\vskip0.2cm

(v) Construction of the Rigged Hilbert Space.

\vskip0.2cm

\noindent Steps (i)-(iv) will be addressed by using the Sturm-Liouville 
theory~\cite{DUNFORD}.  Step (v) will be handled by the methods 
proposed in Ref.~\cite{DIS}.

In Section~\ref{sec:fdosehami}, we construct the self-adjoint Hamiltonian
from the formal Schr\"odinger differential operator and from a domain of the 
Hilbert space.  In Section~\ref{sec:greenfunctions}, we construct the Green 
function and the resolvent of $H$.  In Section~\ref{sec:spectrum}, we compute 
the spectrum of the Hamiltonian.  In Section~\ref{sec:directindeco}, we 
construct the direct integral decomposition of the Hilbert space generated by 
the Hamiltonian.  In Section~\ref{sec:constRHS}, the RHS of the square 
well-barrier Hamiltonian is constructed.

\def\thesection{\arabic{section}}
\section{Formal Differential Operator and Self-Adjoint Hamiltonian}
\def\thesection{\arabic{section}}
\setcounter{equation}{0}
\label{sec:fdosehami}

The first step toward the construction of a RHS associated to a Hamiltonian
is to define that Hamiltonian.  In non-relativistic quantum mechanics,
the Hamiltonian is defined by the formal differential operator of the
time-independent Schr\"odinger equation and by a Hilbert space domain on 
which the Hamiltonian acts.

The time-independent Schr\"odinger equation reads in the position 
representation as
\begin{equation}
       \left(\frac{-\hbar ^2}{2m} \Delta +V(\vec{x})\right)
	\langle \vec{x}|E\rangle =E\langle \vec{x}|E\rangle \, ,
	\label{delta}
\end{equation}
where $\Delta$ is the three-dimensional Laplacian and 
\begin{equation}
	V(\vec{x})\equiv V(r)=\left\{ \begin{array}{cl}
                                -V_1   &0<r<a  \\
                                V_2 &a<r<b  \\
                                0   &b<r<\infty 
                  \end{array} 
                 \right. 
	\label{potential}
\end{equation}
is the well-square potential, $V_1$ and $V_2$ being two positive real 
numbers.  The potential $V(\vec{x})$ is rotational invariant, and spherical 
coordinates $\vec{x}\equiv (r,\theta, \phi)$ can be used to write 
Eq.~(\ref{delta}) as 
\begin{equation}
      \left(\frac{-\hbar ^2}{2m}\frac{1}{r}\frac{\partial ^2}{\partial r^2}r
	+\frac{\hbar ^2l(l+1)}{2mr^2}+V(r)\right)
        \langle r,\theta, \phi|E,l,l_3\rangle  
        = E\langle r,\theta, \phi|E,l,l_3\rangle\,.
	\label{sphSe}
\end{equation}
By splitting radial and angular dependences, 
\begin{equation}
	\langle r,\theta, \phi|E,l,l_3\rangle \equiv
	\langle r|E\rangle _l \, \langle \theta, \phi |l,l_3\rangle \equiv
	\frac{1}{r}\chi _l(r;E)Y_{l,l_3}(\theta, \phi),
\end{equation}
where $Y_{l,l_3}(\theta, \phi)$ are the spherical harmonics, we obtain for 
the radial part
\begin{equation}
	\left(\frac{-\hbar ^2}{2m}\frac{d^2}{dr^2}+\frac{\hbar ^2l(l+1)}{2mr^2}
	+V(r) \right) \chi _l(r;E)=E\chi _l(r;E)\,.
	\label{baba}
\end{equation}
For simplicity we study only the case $l=0$.  Writing 
$\chi _{l=0}(r;E)\equiv \chi (r;E)$,
\begin{equation}
       \left( -\frac{\hbar ^2}{2m} \frac{d^2}{dr^2}+V(r) \right) \chi (r;E)=
        E\chi (r;E)\,.
	\label{rSe0}
\end{equation}
Therefore, the formal Schr\"odinger differential operator for the zero angular 
momentum case is
\begin{equation}
       h \equiv -\frac{\hbar ^2}{2m} \frac{d^2}{dr^2}+V(r) \, .
      \label{doh}
\end{equation}

In order to construct the domain ${\cal D}(H)$ of the Hamiltonian, we need 
first the Hilbert space that contains it.  Clearly, the Hilbert space that
fits the differential operator (\ref{doh}) is $L^2([0,\infty ),dr)$.  Now, 
each element $f$ in the domain ${\cal D}(H)$ must be square integrable (see 
condition (\ref{HSc}) below), the action of $h$ on $f$ must be well-defined 
(see condition (\ref{ACc}) below), and the action of $h$ on $f$ must remain 
square integrable (see condition (\ref{reduce}) below).  Furthermore, the 
Hamiltonian $H$ induced by $h$ and by ${\cal D}(H)$ must be self-adjoint.  It 
is well-known that all the possible self-adjoint operators associated to the 
formal differential operator (\ref{doh}) are determined by the following 
condition (cf.~\cite{DUNFORD}, page 1306):
\begin{equation}
       f(0)+\alpha  \, f'(0)=0 \, ,
       \quad -\infty < \alpha \leq \infty \, .
       \label{sabocu}
\end{equation}
Among all the possibilities in Eq.~(\ref{sabocu}), the one that is used in 
physics is $\alpha =0$, i.e., $f(0)=0$.  All the above mentioned conditions,
which determine the domain ${\cal D}(H)$, can be written as
\begin{mathletters}
\begin{eqnarray}
      f(r) \in L^2([0,\infty ), dr)  \, , \label{HSc} \\
      f(r) \in AC^2 [0,\infty ) \, ,  \label{ACc} \\
      (h f )(r) \in L^2([0,\infty ), dr) \, , \label{reduce} \\
       f(0)=0 \, , \label{sac}  
\end{eqnarray}
\label{bcthdpspr} 
\end{mathletters}
where $AC^2 ([0,\infty ) )$ denotes the space of 
all functions $f$ which have a continuous derivative in 
$[0,\infty )$, and for which $f'$ is not only continuous but also 
absolutely continuous over each compact subinterval of $[0,\infty )$.  Thus
the domain of our Hamiltonian is 
\begin{equation}
{\cal D}(H)=\{ f\in L^2([0,\infty ), dr) \, | \ f\in AC^2 [0,\infty ),
             hf\in L^2([0,\infty ), dr), f(0)=0 \} \, .
          \label{domain}
\end{equation}
It follows that the Hamiltonian defined by
\begin{equation}
       (Hf)(r):=(hf)(r) \, , \quad f\in {\cal D}(H) 
       \label{operatora}
\end{equation}
is a well-defined self-adjoint linear operator.

\def\thesection{\arabic{section}}
\section{Resolvent and Green Function}
\def\thesection{\arabic{section}}
\setcounter{equation}{0}
\label{sec:greenfunctions}

The second step toward the construction of the RHS is to compute the 
resolvent and the Green function of the (self-adjoint) Hamiltonian.  The
resolvent operator of $H$ is defined as the inverse of the operator
$(E-H)$ for those complex energies $E$ for which that inverse exists.  If
we write the resolvent operator as an integral operator, the kernel of
that integral operator is precisely the Green function,
\begin{equation}
      \left[ (E-H)^{-1}f\right] (r)=\int_0^{\infty}ds \, G(r,s;E) f(s) \, .
\end{equation}

The Green function $G(r,s;E)$ can be computed using the prescription provided
by the following theorem (cf.~Theorem~XIII.3.16 of Ref.~\cite{DUNFORD} and
also Ref.~\cite{CSF}):

\vskip0.5cm

{\bf Theorem~1}\quad  Let $H$ be the self-adjoint operator (\ref{operatora}) 
derived from the real formal differential operator (\ref{doh}) by the 
imposition of the boundary condition (\ref{sac}).  Let 
$\mbox{Im}(E)\neq 0$.  Then 
there is exactly one solution $\chi (r;E)$ of $(h-E)\sigma =0$ 
square-integrable at $0$ and satisfying the boundary condition (\ref{sac}), 
and exactly one solution $\Theta (r;E)$ of $(h-E)\sigma =0$ square-integrable 
at infinity.  The resolvent $(E-H)^{-1}$ is an integral operator whose kernel 
$G(r,s;E)$ is given by
\begin{equation}
       G(r,s;E)=\left\{ \begin{array}{ll}
               \frac{2m}{\hbar ^2} \,
      \frac{\chi (r;E) \, \Theta (s;E)}{W(\chi ,\Theta )}
               &r<s \\ 
      \frac{2m}{\hbar ^2} \,
      \frac{\chi (s;E) \, \Theta (r;E)}{W(\chi ,\Theta )}
                       &r>s  \, ,
                  \end{array} 
                 \right. 
	\label{exofGFA}
\end{equation}
where $W(\chi ,\Theta )$ is the Wronskian of $\chi$ and $\Theta$,
\begin{equation}
       W(\chi ,\Theta )=\chi \Theta '-\chi ' \Theta \, .
\end{equation}

\vskip0.5cm

In order to compute $G(r,s;E)$, we shall divide the complex energy plane in 
three different regions and obtain $G(r,s;E)$ for each region separately.  In
our calculations, we shall use the following branch of the square root 
function:
\begin{equation}
      \sqrt{\cdot}:\{ E\in {\mathbb C} \, | \  -\pi <{\rm arg}(E)\leq \pi \} 
   \longmapsto \{E\in {\mathbb C} \, | \  -\pi/2 <{\rm arg}(E)\leq \pi/2 \} 
      \, .
   \label{branch} 
\end{equation}

\def\thesubsection{\thesection.\arabic{subsection}}
\subsection{Region ${\rm Re}(E)<0$, ${\rm Im}(E)\neq 0$}

We first apply Theorem~1 to the region of the complex energy plane where 
${\rm Re}(E)<0$ and ${\rm Im}(E)\neq 0$.  According to that theorem, the 
Green function reads
\begin{equation}
       G(r,s;E)=\left\{ \begin{array}{ll}
               -\frac{2m/\hbar ^2}{\sqrt{-2m/\hbar ^2 \, E}} \,
    \frac{\widetilde{\chi}(r;E) \, \widetilde{\Theta}(s;E)}
          {2\widetilde{\mathcal{J}}_3(E)}
               &r<s \\ 
     -\frac{2m/\hbar ^2}{\sqrt{-2m/\hbar ^2 \, E}} \,
    \frac{\widetilde{\chi}(s;E) \, \widetilde{\Theta}(r;E)}
          {2\widetilde{\mathcal{J}}_3(E)}
                       &r>s  
                  \end{array} 
                 \right. \quad \mbox{Re}(E)<0 \, , \ \mbox{Im}(E)\neq 0 \, .
      \label{green-}
\end{equation}
In this equation, the eigenfunction $\widetilde{\chi}(r;E)$ satisfies the
time-independent Schr\"odinger equation (\ref{rSe0}) subject to the following 
boundary conditions:
\begin{mathletters}
\begin{eqnarray}
       && \widetilde{\chi} (0;E)=0 \, , \label{bca01} \\
       && \widetilde{\chi} (r;E)\in AC^2([0,\infty )) \, , \label{bcchiac} \\
       && \widetilde{\chi}(r;E)
           {\rm \ is \ square \ integrable \ at \ } 0 \, ,
       \label{sbca03} 
\end{eqnarray}
      \label{eigsbo0co}
\end{mathletters}
which yield
\begin{equation}
      \widetilde{\chi}(r;E)=
          \left\{ \begin{array}{lll}
             \frac{i}{2}\, (e^{\sqrt{-\frac{2m}{\hbar ^2}(E+V_1)}\, r}-
            e^{-\sqrt{-\frac{2m}{\hbar ^2}(E+V_1)}\, r})
                 \quad &0<r<a  \\
            \widetilde{\cal J}_1(E)e^{\sqrt{- \frac{2m}{\hbar ^2}(E-V_2)}\, r}
           +\widetilde{\cal J}_2(E)e^{-\sqrt{- \frac{2m}{\hbar ^2}(E-V_2)}\, r}
                 \quad  &a<r<b \\
            \widetilde{{\cal J}}_3(E)e^{\sqrt{- \frac{2m}{\hbar ^2}E}\, r}
            +\widetilde{{\cal J}}_4(E)e^{-\sqrt{- \frac{2m}{\hbar ^2}E}\, r}
                 \quad  &b<r<\infty \, .
               \end{array} 
                 \right. 
       \label{tildechifunction}
\end{equation}
The expressions of the functions 
$\widetilde{\cal J}_1(E)$-$\widetilde{\cal J}_4(E)$ can be found in 
Appendix~\ref{sec:appendix}.  

The eigenfunction $\widetilde{\Theta}(r;E)$ of Eq.~(\ref{green-}) satisfies 
the time-independent Schr\"odinger equation (\ref{rSe0}) subject to the 
following boundary conditions:
\begin{mathletters}
\begin{eqnarray}
       &&\widetilde{\Theta}(r;E)\in AC^2([0,\infty )) \, , \label{bcainfty1} \\
       &&\widetilde{\Theta} (r;E) \ 
         {\rm is \ square \ integrable \ at \ } \infty \, ,
\end{eqnarray}
      \label{thetcoabejej}
\end{mathletters}
which yield
\begin{equation}
      \widetilde{\Theta}(r;E)=
      \left\{ \begin{array}{lll}
         \widetilde{{\cal A}}_1(E)e^{\sqrt{-\frac{2m}{\hbar ^2}(E+V_1)}\, r}
         +\widetilde{{\cal A}}_2(E)e^{-\sqrt{-\frac{2m}{\hbar ^2}(E+V_1)}\, r} 
           \quad &0<r<a  \\
         \widetilde{{\cal A}}_3(E)e^{\sqrt{-\frac{2m}{\hbar ^2}(E-V_2)}\, r}
         +\widetilde{{\cal A}}_4(E)e^{-\sqrt{-\frac{2m}{\hbar ^2}(E-V_2)}\, r}
                 \quad  &a<r<b \\
               e^{-\sqrt{-\frac{2m}{\hbar ^2}E}\, r}
                 \quad  &b<r<\infty \, . 
               \end{array} 
                 \right. 
      \label{tildethetfunc}
\end{equation}
The expressions of the functions 
$\widetilde{\cal A}_1(E)$-$\widetilde{\cal A}_4(E)$ can be found in 
Appendix~\ref{sec:appendix}.

\subsection{Region ${\rm Re}(E)>0$, ${\rm Im}(E)> 0$}
\label{sec:region++}

In this region of the complex plane, Theorem~1 leads to the following Green 
function:
\begin{equation}
       G(r,s;E)=\left\{ \begin{array}{ll}
               \frac{2m/\hbar ^2}{\sqrt{2m/\hbar ^2 \, E}} \,
      \frac{\chi (r;E) \, \Theta _+(s;E)}{2i{\cal J}_4(E)}
               &r<s \\ 
      \frac{2m/\hbar ^2}{\sqrt{2m/\hbar ^2 \, E}} \,
      \frac{\chi (s;E) \, \Theta _+(r;E)}{2i{\cal J}_4(E)}
                       &r>s  
                  \end{array} 
                 \right.  \quad {\rm Re}(E)>0 \, , \ {\rm Im}(E)> 0 \, .
      \label{green++} 
\end{equation}
In this equation, the eigenfunction $\chi (r;E)$ satisfies the differential 
equation (\ref{rSe0}) subject to the boundary conditions (\ref{eigsbo0co}),
\begin{equation}
      \chi (r;E)=\left\{ \begin{array}{lll}
               \sin (\sqrt{\frac{2m}{\hbar ^2}(E+V_1)}\, r) \quad &0<r<a  \\
               {\cal J}_1(E)e^{i \sqrt{\frac{2m}{\hbar ^2}(E-V_2)}\, r}
                +{\cal J}_2(E)e^{-i\sqrt{\frac{2m}{\hbar ^2}(E-V_2)}\, r}
                 \quad  &a<r<b \\
               {\cal J}_3(E) e^{i\sqrt{\frac{2m}{\hbar ^2}E}\, r}
                +{\cal J}_4(E)e^{-i\sqrt{\frac{2m}{\hbar ^2}E}\, r}
                 \quad  &b<r<\infty \, .
               \end{array} 
                 \right. 
             \label{chi}
\end{equation}
The functions ${\cal J}_1(E)$-${\cal J}_4(E)$ are determined by the boundary 
conditions (\ref{eigsbo0co}), and their expressions are listed in
Appendix~\ref{sec:appendix}. 

The eigenfunction $\Theta _+ (r;E)$ satisfies the differential equation
(\ref{rSe0}) subject to the boundary conditions (\ref{thetcoabejej}),
\begin{equation}
      \Theta _+(r;E)=\left\{ \begin{array}{lll}
               {\cal A}^+_1(E)e^{i\sqrt{\frac{2m}{\hbar ^2}(E+V_1)}\, r}
               +{\cal A}^+_2(E) e^{-i\sqrt{\frac{2m}{\hbar ^2}(E+V_1)}\, r} 
                \quad &0<r<a  \\
               {\cal A}^+_3(E)e^{i\sqrt{\frac{2m}{\hbar ^2}(E-V_2)}\, r}
                +{\cal A}^+_4(E)e^{-i\sqrt{\frac{2m}{\hbar ^2}(E-V_2)}\, r}
                 \quad  &a<r<b \\
               e^{i\sqrt{\frac{2m}{\hbar ^2}E}\, r}
                 \quad  &b<r<\infty \, . 
               \end{array} 
                 \right. 
            \label{theta+fun}
\end{equation}
The functions ${\cal A}^+_1(E)$-${\cal A}^+_4(E)$ are determined by the boundary
conditions (\ref{thetcoabejej}), and their expressions are listed in
Appendix~\ref{sec:appendix}. 

\subsection{Region ${\rm Re}(E)>0$, ${\rm Im}(E)< 0$}

If ${\rm Re}(E)>0$, ${\rm Im}(E)< 0$ and we use the branch of the square 
root function~(\ref{branch}), then the Green function has the form
\begin{equation}
       G(r,s;E)=\left\{ \begin{array}{ll}
               -\frac{2m/\hbar ^2}{\sqrt{2m/\hbar ^2\, E}} \,
      \frac{\chi (r;E) \, \Theta _-(s;E)}{2i{\cal J}_3(E)}
               &r<s \\ 
      -\frac{2m/\hbar ^2}{\sqrt{2m/\hbar ^2 \, E}} \,
      \frac{\chi (s;E) \, \Theta _-(r;E)}{2i{\cal J}_3(E)}
                       &r>s  
                  \end{array} 
                 \right. \quad {\rm Re}(E)>0 \, , \ {\rm Im}(E)< 0 \, . 
        \label{green+-}
\end{equation}
The eigenfunction $\chi (r;E)$ satisfies the time-independent Schr\"odinger 
equation (\ref{rSe0}) subject to the boundary conditions 
(\ref{eigsbo0co}).  Thus $\chi (r;E)$ is given by Eq.~(\ref{chi})---hence
the same symbol as in the previous subsection---although now $E$ lies in 
the fourth quadrant of the complex energy plane. 

The eigenfunction $\Theta _-(r;E)$ satisfies the differential equation
(\ref{rSe0}) subject to the boundary conditions (\ref{thetcoabejej}),
\begin{equation}
      \Theta _-(r;E)=\left\{ \begin{array}{lll}
               {\cal A}^-_1(E)e^{i\sqrt{\frac{2m}{\hbar ^2}(E+V_1)}\, r}
               +{\cal A}^-_2(E) e^{-i\sqrt{\frac{2m}{\hbar ^2}(E+V_1)}\, r} 
                \quad &0<r<a  \\
               {\cal A}^-_3(E)e^{i\sqrt{\frac{2m}{\hbar ^2}(E-V_2)}\, r}
                +{\cal A}^-_4(E)e^{-i\sqrt{\frac{2m}{\hbar ^2}(E-V_2)}\, r}
                 \quad  &a<r<b \\
               e^{-i\sqrt{\frac{2m}{\hbar ^2}E}\, r}
                 \quad  &b<r<\infty \, . 
               \end{array} 
                 \right. 
            \label{thetafun-}
\end{equation}
The functions ${\cal A}^-_1(E)$-${\cal A}^-_4(E)$ are determined by the 
boundary conditions (\ref{thetcoabejej}), and their expressions are listed in
Appendix~\ref{sec:appendix}.

\def\thesection{\arabic{section}}
\section{Spectrum of the Hamiltonian}
\def\thesection{\arabic{section}}
\setcounter{equation}{0}
\label{sec:spectrum}

The third step toward the construction of the RHS is to obtain the (Hilbert
space) spectrum $\mbox{Sp}(H)$ of the self-adjoint Hamiltonian $H$.  Since
$H$ is self-adjoint, the spectrum must be real.  In order to elucidate which
real numbers are in $\mbox{Sp}(H)$ we make use of Theorem~3 below.  Before
stating Theorem~3, we need to state another theorem, which provides the 
unitary operator $U$ that will be used to diagonalize $H$ 
(cf.~Theorem XIII.5.13 of Ref.~\cite{DUNFORD}):

\vskip0.5cm

{\bf Theorem~2} (Weyl-Kodaira)\quad  Let $h$ be the formally self-adjoint 
differential operator (\ref{doh}) defined on the interval 
$[0,\infty )$.  Let $H$ be the self-adjoint operator (\ref{operatora}).  Let
$\Lambda$ be an open interval of the real axis, and suppose that there 
is given a set $\{ \sigma _1(r;E),\, \sigma _2(r;E)\}$ of functions, defined 
and continuous on $(0,\infty )\times \Lambda$, such that for each fixed 
$E$ in $\Lambda$, $\{ \sigma _1(r;E),\, \sigma _2(r;E)\}$ forms a basis for
the space of solutions of $h\sigma =E\sigma$.  Then there exists a 
positive $2\times 2$ matrix measure $\{ \rho _{ij} \}$ defined on
$\Lambda$, such that the limit 
\begin{equation}
      (U f)_i(E):=\lim_{c\to 0}\lim_{d\to \infty} 
        \left[ \int_c^d f(r) \overline{\sigma _i(r;E)}dr \right]
\end{equation}
exists in the topology of $L^2(\Lambda ,\{ \rho _{ij}\})$ for each 
$f$ in $L^2([0,\infty ),dr)$ and defines an isometric isomorphism $U$ of
${\sf E}(\Lambda )L^2([0,\infty ),dr)$ onto $L^2(\Lambda ,\{ \rho _{ij}\})$, 
${\sf E}(\Lambda )$ being the spectral projection associated to $\Lambda$.
      
\vskip0.5cm

The spectral measures $\{ \rho _{ij}\}$ are provided by the following theorem 
(cf.~Theorem XIII.5.18 of Ref.~\cite{DUNFORD}):

\vskip0.5cm

{\bf Theorem~3} (Titchmarsh-Kodaira)\quad  Let $\Lambda$ be an open 
interval of the real axis, $O$ be an open set in the complex plane 
containing $\Lambda$, and ${\rm Re}(H):={\mathbb C}-{\rm Sp}(H)$.  Let 
$\{ \sigma _1(r;E),\, \sigma _2(r;E)\}$ be a set of functions 
which form a basis for the solutions of the equation $h\sigma =E\sigma$, 
$E\in O$, and which are continuous on $(0,\infty )\times O$ and analytically 
dependent on $E$ for $E$ in $O$.  Suppose that the kernel $G(r,s;E)$ for the 
resolvent $(E-H)^{-1}$ has a representation
\begin{equation}
      G(r,s;E)=\left\{ \begin{array}{lll}
                   \sum_{i,j=1}^2 \theta _{ij}^-(E)\sigma _i(r;E)
                   \overline{\sigma _j(s;\overline{E})} &
                     \qquad & r<s   \\
                  \sum_{i,j=1}^2 \theta _{ij}^+(E)\sigma _i(r;E)
                   \overline{\sigma _j(s;\overline{E})} &\qquad & r>s \, ,
                  \end{array}
                 \right.
       \label{greenfunitthes}
\end{equation}
for all $E$ in ${\rm Re}(H)\cap O$, and that $\{ \rho _{ij} \}$ is a 
positive matrix measure on $\Lambda$ associated with $H$ as in Theorem 2.  Then
the functions $\theta _{ij}^{\pm}$ are analytic in ${\rm Re}(H)\cap O$, and
given any bounded open interval $(E_1,E_2)\subset \Lambda$, we have for 
$1\leq i,j\leq 2$,
\begin{equation}
       \begin{array}{lll}
       \rho _{ij}((E_1,E_2))&=& \lim_{\delta \to 0}\lim_{\epsilon \to 0+}
         \frac{1}{2\pi i}\int_{E_1+\delta}^{E_2-\delta}
          [ \theta _{ij}^-(E-i\epsilon )-\theta _{ij}^-(E+i\epsilon )
          ]dE \\ [2ex]
      \quad &=& \lim_{\delta \to 0}\lim_{\epsilon \to 0+}
         \frac{1}{2\pi i}\int_{E_1+\delta}^{E_2-\delta}
          [ \theta _{ij}^+(E-i\epsilon )-\theta _{ij}^+(E+i\epsilon )
          ]dE \, .
         \end{array} 
\end{equation}

\vskip0.5cm

Using these spectral measures we can compute the spectrum of $H$.  This 
spectrum is the subset of the real line on which the Green function fails to 
be analytic.  This non-analyticity of $G(r,s;E)$ will be built into the 
functions $\theta _{ij}^{\pm}(E)$ that appear in Theorem~3.  From the 
expression of the Green function computed in Section~\ref{sec:greenfunctions},
it is clear that the subsets $(-\infty ,0)$ and $(0,\infty )$ should be 
studied separately.  We will denote either of these subsets by $\Lambda$.

\subsection{Negative Energy Real Line: $\Lambda =(-\infty ,0)$}
\label{sec:NeERlin}

We first take $\Lambda$ from Theorem~3 to be $(-\infty ,0)$.  We choose a 
basis for the space of solutions of the equation $h\sigma =E\sigma$ that is 
continuous on $(0,\infty )\times \Lambda$ and analytically dependent on $E$ as
\begin{mathletters}
\begin{eqnarray}
      &&\sigma _1(r;E)=\left\{ \begin{array}{lll}
               \widetilde{\cal B}_1(E)e^{\sqrt{-\frac{2m}{\hbar ^2}(E+V_1)}r}
                +\widetilde{\cal B}_2(E)
                  e^{-\sqrt{-\frac{2m}{\hbar ^2}(E+V_1)}r} 
                \quad &0<r<a  \\
               \widetilde{\cal B}_3(E)e^{\sqrt{-\frac{2m}{\hbar ^2}(E-V_2)}r}
                +\widetilde{\cal B}_4(E)
                 e^{-\sqrt{-\frac{2m}{\hbar ^2}(E-V_2)}r}
                 \quad  &a<r<b \\
               e^{\sqrt{-\frac{2m}{\hbar ^2}E}r}
                 \quad  &b<r<\infty   \, ,
               \end{array} 
                 \right. \qquad  \label{tildesigma1} \\  
      &&\sigma _2(r;E)=\widetilde{\Theta}(r;E) \, .
\end{eqnarray}
\end{mathletters}
The functions $\widetilde{\cal B}_1$-$\widetilde{\cal B}_4$ are such that
$\sigma _1(r;E)$ and its derivative are continuous at $r=a$ and
at $r=b$.  Their expressions are listed in Eq.~(\ref{tildeBfunctions}) of
Appendix~\ref{sec:appendix}.  The function $\widetilde{\Theta}(r;E)$ is 
given by Eq.~(\ref{tildethetfunc}).  Obviously,
\begin{equation}
      \widetilde{\chi}(r;E)=\widetilde{{\cal J}}_3(E)\sigma _1(r;E)+
      \widetilde{{\cal J}}_4(E) \sigma _2(r;E) \, ,
\end{equation}
which along with Eq.~(\ref{green-}) leads to
\begin{eqnarray}
      &&G(r,s;E)= 
      -\frac{2m/\hbar ^2}{\sqrt{-2m/\hbar ^2 \, E}} \,
       \frac{1}{2}\, \left[ \sigma _1(r;E)
       +\frac{\widetilde{{\cal J}}_4(E)}{\widetilde{{\cal J}}_3(E)} \, 
       \sigma _2(r;E)\right] \sigma _2(s;E) \, , 
       \nonumber \\
        &&\quad \hskip7cm r<s \, , \  \mbox{Re}(E)<0 \, , \mbox{Im}(E) 
            \neq 0 \, . 
        \label{dis-parosof}
\end{eqnarray}
Since
\begin{equation}
      \overline{\sigma _2(s;\overline{E})}=
      \sigma _2(s;E) \, ,
\end{equation}
we can write Eq.~(\ref{dis-parosof}) as
\begin{eqnarray}
       &&G(r,s;E)= 
      -\frac{2m/\hbar ^2}{\sqrt{-2m/\hbar ^2 \, E}} \,
       \frac{1}{2}\,  \left[ \sigma _1(r;E)
       \overline{\sigma _2(s;\overline{E})}
       +\frac{\widetilde{{\cal J}}_4(E)}{\widetilde{{\cal J}}_3(E)} \, 
       \sigma _2(r;E)
       \overline{\sigma _2(s;\overline{E})} \right] , \nonumber \\
       &&\qquad \hskip7cm
        r<s \, , \  \mbox{Re}(E)<0 \, , \mbox{Im}(E) \neq 0 \, .
        \label{finag-retoco} 
\end{eqnarray}
By comparing Eqs.~(\ref{greenfunitthes}) and (\ref{finag-retoco}) we see that
\begin{equation}
      \theta _{ij}^-(E)= \left(  \begin{array}{cc}
        0 & -\frac{2m/\hbar ^2}{\sqrt{-2m/\hbar ^2 \, E}} 
       \frac{1}{2}  \\
        0 &
      -\frac{2m/\hbar ^2}{\sqrt{-2m/\hbar ^2 \, E}} 
       \frac{1}{2}  
       \frac{\widetilde{\cal J}_4(E)}{\widetilde{\cal J}_3(E)} 
                                \end{array}
        \right) \, , \quad \mbox{Re}(E)<0 \, , \  \mbox{Im}(E) \neq 0 \, .
\end{equation}
The functions $\theta _{ij}^-(E)$ are analytic in a neighborhood of 
$\Lambda =(-\infty ,0)$ except at the energies for which 
$\widetilde{\cal J}_3(E)$ vanishes.  In this case, the function 
$\theta _{22}^-(E)$ has a pole.  Since
\begin{equation}
      \widetilde{\cal J}_3(\tilde{k})=\widetilde{\cal J}_3(-ik)=
      {\cal J}_4(k)=\frac{i}{2}\,{\cal J}_+(k) \, , 
\end{equation}
where ${\cal J}_+(k)$ is the Jost function~\cite{HOGREVE}, the poles of
$\theta _{22}^-(E)$ correspond to the zeros of the Jost function on the
negative real axis of the energy plane, or on the positive imaginary 
$k$-axis.  Those bound states have been calculated by Hogreve~\cite{HOGREVE} 
for different values of $V_1$, $V_2$ in Eq.~(\ref{potential}).  In general,
there is a finite number of bound states 
whose energies $E_1,\ldots ,E_N$ are in the interval $(-V_1,0)$.  The 
corresponding normalized wave functions read~\cite{DUNFORD}
\begin{equation}
      \phi _n(r)=N_n\, \widetilde{\Theta}(r;E_n) \, , 
       \quad n=1,\ldots ,N \, ,
       \label{normaiegin} 
\end{equation}
where the square of $N_n$ is given by the residue of the spectral function 
$\theta _{22}^-(E)$ at the bound state energy $E_n$,
\begin{equation}
     N_n^2=\mbox{res}\left[\theta _{22}^-(E)\right]_{E=E_n}=
       \mbox{res}\left[-\frac{2m/\hbar ^2}{\sqrt{-2m/\hbar ^2 \, E}} 
       \frac{1}{2}  
       \frac{\widetilde{\cal J}_4(E)}{\widetilde{\cal J}_3(E)}
        \right]_{E=E_n}\, .
       \label{normalizationi}
\end{equation}
Therefore $E_1,\ldots ,E_N$ are the only negative energies that belong
to the spectrum of our Hamiltonian.

It is worthwhile noting that the normalization (\ref{normalizationi}) 
provided by the residue of the spectral measure $\theta _{22}^-(E)$ is the
same (as it should!) as the normalization obtained in 
Refs.~\cite{NEWTON,TAYLOR,NUSSENZVEIG}.  To see this, we denote the momentum 
of those bound states by $k_{n}=i|k_{n}|=i\widetilde{k}_n$, 
$n=1,\ldots ,N$.  Then we have that
\begin{equation}
      \widetilde{\Theta}(r;\widetilde{k}_n)=\widetilde{\Theta}(r;-i{k}_n)=
      \Theta _+(r;{k}_n) \, ,
\end{equation}
and that
\begin{equation}
      \mbox{res}\left[-\frac{2m/\hbar ^2}{\sqrt{-2m/\hbar ^2 \, E}} 
       \frac{1}{2}  
       \frac{\widetilde{\cal J}_4(E)}{\widetilde{\cal J}_3(E)}
        \right]_{E=E_n}=
        \mbox{res}\left[i\frac{2m/\hbar ^2}{\sqrt{2m/\hbar ^2 \, E}} 
       \frac{1}{2}  
       \frac{{\cal J}_-(E)}{{\cal J}_+(E)}\right]_{E=E_n} =
        i\, \mbox{res}\left[S(k) \right]_{k=k_n} \, , 
       \label{normalizationiaa}
\end{equation}
where $S(k)$ is the $S$-matrix in the $k$-momentum representation and
\begin{equation}
      {\cal J}_-(E)=2i{\cal J}_3(E) \, . 
\end{equation}
Since the eigenfunction $\phi _n(r)$ of Eq.~(\ref{normaiegin}) is normalized 
to one, from Eqs.~(\ref{normalizationi})-(\ref{normalizationiaa}) it follows 
that
\begin{equation}
      \int_0^{\infty}dr \, \left[\Theta _+(r;k_n)\right]^2 = 
      \frac{-i}{\mbox{res}\left[S(k) \right]_{k=k_n}} \, ,
      \label{normanewtoon}
\end{equation}
which is the normalization that appears in 
Refs.~\cite{NEWTON,TAYLOR,NUSSENZVEIG}.  (The normalization rule of 
Eq.~(\ref{normanewtoon}) has been generalized to Gamow vectors in
Refs.~\cite{MONDRAGON1,MONDRAGON2,MONDRAGON3}.)

\subsection{Positive Energy Real Line: $\Lambda =(0, \infty )$}
\label{sec:PeERlin}

Now we study the case $\Lambda =(0,\infty )$.  In order to be able to 
apply Theorem~3, we choose the following basis
for the space of solutions of $h\sigma =E\sigma$ that is continuous
on $(0,\infty )\times \Lambda$ and analytically dependent on $E$:  
\begin{mathletters}
\begin{eqnarray}
      &&\sigma _1(r;E)=\chi (r;E)\, ,
      \label{sigma1=sci} \\ 
      &&\sigma _2(r;E)=\left\{ \begin{array}{lll}
               \cos (\sqrt{\frac{2m}{\hbar ^2}(E+V_1)}r) \quad &0<r<a  \\
               {\cal C}_1(E)e^{i \sqrt{\frac{2m}{\hbar ^2}(E-V_2)}r}
                +{\cal C}_2(E)e^{-i\sqrt{\frac{2m}{\hbar ^2}(E-V_2)}r}
                 \quad  &a<r<b \\
               {\cal C}_3(E)e^{i\sqrt{\frac{2m}{\hbar ^2}E}r}
                 +{\cal C}_4(E)e^{-i\sqrt{\frac{2m}{\hbar ^2}E}r}
                 \quad  &b<r<\infty \, . 
               \end{array} 
                 \right. \quad
       \label{sigam2cos}  
\end{eqnarray}
    \label{basisin0infisi} 
\end{mathletters}
The functions ${\cal C}_1$-${\cal C}_4$, whose expressions are given by
Eq.~(\ref{Cfunctions}) of Appendix~\ref{sec:appendix}, are such that 
$\sigma _2$ and its derivative are continuous at $r=a$ and at $r=b$.  The eigenfunction
$\chi (r;E)$ is given by Eq.~(\ref{chi}).  

Eqs.~(\ref{theta+fun}), (\ref{thetafun-}) and (\ref{basisin0infisi}) lead to  
\begin{mathletters}
\begin{eqnarray}
      &&\Theta _+(r;E)=-\frac{{\cal C}_4(E)}{W(E)} \sigma _1(r;E)+
      \frac{{\cal J}_4(E)}{W(E)}\sigma _2(r;E) \, ,
      \label{Thet+inosigm} \\
      &&\Theta _-(r;E)=\frac{{\cal C}_3(E)}{W(E)} \sigma _1(r;E)-
      \frac{{\cal J}_3(E)}{W(E)}\sigma _2(r;E) \, ,
      \label{Thet-inosigm}
\end{eqnarray}
\end{mathletters}
where
\begin{equation}
       W(E)={\cal J}_4(E){\cal C}_3(E)-{\cal J}_3(E){\cal C}_4(E) \, .
\end{equation}
By substituting Eq.~(\ref{Thet+inosigm}) into Eq.~(\ref{green++}) we get to
\begin{eqnarray}
      &&G(r,s;E)=
       \frac{2m/\hbar ^2}{\sqrt{2m/\hbar ^2 \, E}} \,
       \frac{1}{2i{\cal J}_4(E)}\,
       \left[ -\frac{ {\cal C}_4(E)}{W(E)}
       \sigma _1(r;E)+\frac{ {\cal J}_4(E)}{W(E)}\sigma _2(r;E)\right] 
       \sigma _1 (s;E) \, ,  \nonumber \\
      &&\qquad \hskip7cm  \mbox{Re}(E)>0, \mbox{Im}(E)>0\, , \, r>s \, .
        \label{G++tofpuon}
\end{eqnarray}
By substituting Eq.~(\ref{Thet-inosigm}) into
Eq.~(\ref{green+-}) we get to
\begin{eqnarray}
      &&G(r,s;E)=
      -\frac{2m/\hbar ^2}{\sqrt{2m/\hbar ^2 \, E}}   \,
       \frac{1}{2i{\cal J}_3(E)}\,
       \left[ \frac{ {\cal C}_3(E)}{W(E)}
       \sigma _1(r;E)-\frac{{\cal J}_3(E)}{W(E)}\sigma _2(r;E)\right]      
       \sigma _1 (s;E) \, , \nonumber \\
      &&\qquad \hskip7cm \mbox{Re}(E)>0, \mbox{Im}(E)<0\, , \, r>s \, .
      \label{G+-tofpuon}
\end{eqnarray} 
Since
\begin{equation}
      \overline{\sigma _1(s;\overline{E})}=\sigma _1(s;E) \, ,
\end{equation}
Eq.~(\ref{G++tofpuon}) can be written as
\begin{eqnarray}
      &&G(r,s;E)=
       \frac{2m/\hbar ^2}{\sqrt{2m/\hbar ^2 \, E}} \,
       \frac{1}{2i{\cal J}_4(E)}\,
       \left[ -\frac{ {\cal C}_4(E)}{W(E)}
       \sigma _1(r;E)\overline{\sigma _1(s;\overline{E})}   
      +\frac{ {\cal J}_4(E)}{W(E)}\sigma _2(r;E)
       \overline{\sigma _1(s;\overline{E})}\right] \nonumber \\      
      &&\qquad \hskip8cm \mbox{Re}(E)>0, \mbox{Im}(E)>0\, , \, r>s \, ,
       \label{redaot++}
\end{eqnarray}
whereas Eq.~(\ref{G+-tofpuon}) can be written as 
\begin{eqnarray}
      &&G(r,s;E)=
      -\frac{2m/\hbar ^2}{\sqrt{2m/\hbar ^2 \, E}} \,
       \frac{1}{2i{\cal J}_3(E)}\,
       \left[ \frac{ {\cal C}_3(E)}{W(E)}
       \sigma _1(r;E)\overline{\sigma _1(s;\overline{E})}
       -\frac{{\cal J}_3(E)}{W(E)}\sigma _2(r;E)
       \overline{\sigma _1(s;\overline{E})}\right] \nonumber \\
       &&\qquad \hskip8cm \mbox{Re}(E)>0, \mbox{Im}(E)<0\, , \, r>s \, .
       \label{redaot+-}
\end{eqnarray}
By comparing (\ref{greenfunitthes}) to (\ref{redaot++}) we get to
\begin{equation}
      \theta _{ij}^+(E)= \left(  \begin{array}{cc}
      \frac{2m/\hbar ^2}{\sqrt{2m/\hbar ^2 \, E}}
      \frac{1}{2i} \frac{- {\cal C}_4(E)}{ {\cal J}_4(E)W(E)}
       & 0 \\
      \frac{2m/\hbar ^2}{\sqrt{2m/\hbar ^2 \, E}}
      \frac{1}{2i} \frac{1}{W(E)}
       & 0
                                \end{array}
        \right) 
       \, , \quad  \mbox{Re}(E)>0 \, , \  \mbox{Im}(E)>0 \, .
       \label{theta++}
\end{equation}
By comparing (\ref{greenfunitthes}) to (\ref{redaot+-}) we get to
\begin{equation}
      \theta _{ij}^+(E)= \left(  \begin{array}{cc}
      -\frac{2m/\hbar ^2}{\sqrt{2m/\hbar ^2 \, E}}
      \frac{1}{2i} \frac{{\cal C}_3(E)}{ {\cal J}_3(E)W(E)}
       & 0 \\
      \frac{2m/\hbar ^2}{\sqrt{2m/\hbar ^2 \, E}}
      \frac{1}{2i} \frac{1}{W(E)}
       & 0
                                \end{array}
        \right) 
       \, , \quad  \mbox{Re}(E)>0 \, , \  \mbox{Im}(E)<0 \, .
      \label{theta+-}
\end{equation}
From Eqs.~(\ref{theta++}) and (\ref{theta+-}) we can see that the measures 
$\rho _{12}$, $\rho _{21}$ and $\rho _{22}$ in Theorem~3 are zero and that 
the measure $\rho _{11}$ is given by
\begin{eqnarray}
       \rho _{11}((E_1,E_2))&=&\lim _{\delta \to 0} \lim _{\epsilon \to 0+}
      \frac{1}{2\pi i} \int_{E_1+\delta}^{E_2-\delta}
      \left[ \theta _{11}^+ (E-i\epsilon ) -\theta _{11}^+ (E+i\epsilon )
      \right] dE \nonumber \\
      &=&\int_{E_1}^{E_2}   \frac{1}{4\pi}\, 
      \frac{2m/\hbar ^2}{\sqrt{2m/\hbar ^2 \, E}}
      \, \frac{1}{ {\cal J}_3(E) {\cal J}_4(E)}\, dE \, ,
\end{eqnarray}
which leads to
\begin{equation}
      \rho (E)\equiv \rho _{11}(E)=
      \frac{1}{4\pi}\, 
      \frac{2m/\hbar ^2}{\sqrt{2m/\hbar ^2 \, E}}
      \, \frac{1}{ |{\cal J}_4(E)|^2}\, , \quad E\in (0,\infty )  \, .
\end{equation} 
The function $\theta _{11}^+(E)$ has a branch cut along $(0,\infty)$, and 
therefore $(0,\infty )$ is included in ${\rm Sp}(H)$.  Since ${\rm Sp}(H)$ is 
a closed set, we have that
\begin{equation}
      \mbox{Sp}(H)=\{ E_1,\ldots ,E_N \} \cup [0,\infty )\, .
\end{equation}
Therefore, the spectrum of $H$ has a discrete part 
$\Lambda _{\rm b}=\{ E_1,\ldots ,E_N \}$ and a continuous part 
$\Lambda _{\rm c}=[0,\infty )$.

\def\thesection{\arabic{section}}
\section{Direct Integral Decomposition}
\def\thesection{\arabic{section}}
\setcounter{equation}{0}
\label{sec:directindeco}

The fourth step toward the construction of the RHS is to compute the
unitary operator that diagonalizes the Hamiltonian and the direct integral
decomposition induced by that unitary operator.  In order to compute them,
we apply Theorem~2 of Section~\ref{sec:spectrum} to the discrete
and continuous spectrum separately, since they can be treated 
independently.  In the end, the independence of the discrete and continuous
spectra will lead to the splitting of the Hilbert space into a direct sum of 
a Hilbert space associated to the discrete spectrum and another Hilbert space
associated to the continuous spectrum.

We first apply Theorem~2 to the discrete part of the spectrum.  By that 
theorem, there is a unitary operator $U_{\rm b}$ from 
${\mathcal H}_{\rm b} :={\sf E}(\Lambda _{\rm b})L^2([0,\infty ),dr)$ onto
$\widehat{\mathcal{H}}_{\rm b}:={\mathbb C}^N$ defined by
\begin{eqnarray}
     U_{\rm b}: \mathcal{H}_{\rm b}
     &\longmapsto & \widehat{\mathcal{H}}_{\rm b}  \nonumber \\
     f_{\rm b}(r)& \longmapsto & 
     U_{\rm b}f_{\rm b}=\{ 
      (\phi _1,f_{\rm b}), \ldots ,(\phi _N,f_{\rm b}) \} \, .
      \label{Ubound}
\end{eqnarray}
In this equation, $\mathcal{H}_{\rm b}$ is the Hilbert space spanned by the
bound states $\phi _1,\ldots ,\phi _N$, $\widehat{\mathcal{H}}_{\rm b}$ is the
space of $N$-tuples of the form $\{ (\phi _n,f_{\rm b})\}_{n=1}^N$ (hence 
$\widehat{\mathcal{H}}_{\rm b}$ is isomorphic to ${\mathbb C}^N$), 
$f_{\rm b}={\sf E}(\Lambda _{\rm b})f$, i.e., $f_{\rm b}$ is the
component of $f$ along the space $\mathcal{H}_{\rm b}$,
\begin{equation}
      f_{\rm b}(r)=\sum_{n=1}^N (\phi _n,f)\, \phi _n(r) \, ,
\end{equation}
and $(\phi _n,f_{\rm b})$ is the scalar product of the $n$-th bound state
$\phi _n(r)$ of Eq.~(\ref{normaiegin}) with $f_{\rm b}$,
\begin{equation}
      (\phi _n, f_{\rm b})=\int_0^{\infty}dr\, 
       \overline{\phi _n(r)}f_{\rm b}(r) \, , \quad n=1, \ldots , N \, .
\end{equation}
On the space $\widehat{\mathcal{H}}_{\rm b}$, the Hamiltonian acts as the 
following $N\times N$ diagonal matrix:
\begin{equation}
      \widehat{H}_{\rm b}:=U_{\rm b}H_{\rm b}U_{\rm b}^{-1}=
      \left( \begin{array}{cccc} E_1 & 0 & .\, . & 0 \\
                                  0  &E_2 & . \, . & 0 \\           
                                . \, . & . \, . & . \, . & . \, . \\
                                0 & 0 & . \, . & E_N  
             \end{array}         \right) \, , 
       \label{diagonaHb}
\end{equation}
where $H_{\rm b}$ is the restriction of $H$ to $\mathcal{H}_{\rm b}$.

Next, we apply Theorem~2 to the continuous part of the spectrum.  By that
theorem, there is a unitary map $\widetilde{U}_{\rm c}$ from
${\mathcal H}_{\rm c} :={\sf E}(\Lambda _{\rm c})L^2([0,\infty ),dr)$ onto
$L^2( (0,\infty ),\rho (E)dE)$ defined by
\begin{eqnarray}
     \widetilde{U}_{\rm c}:\mathcal{H}_{\rm c} 
      &\longmapsto & L^2( (0,\infty ),\rho (E)dE) \nonumber \\
       f_{\rm c}(r)& \longmapsto & 
       \widetilde{f}_{\rm c}(E)=(\widetilde{U}_{\rm c}f_{\rm c})(E)=
      \int_0^{\infty}dr f_{\rm c}(r) \overline{\chi (r;E)} \, ,
      \label{rhoU}
\end{eqnarray}
where $f_{\rm c}={\sf E}(\Lambda _{\rm c})f$, i.e., $f_{\rm c}$ is the
component of $f$ along the Hilbert space $\mathcal{H}_{\rm c}$.  The space
$\mathcal{H}_{\rm c}$ is the Hilbert space that corresponds to the continuous
part of the spectrum---hence the subscript ${\rm c}$.

If we write
\begin{equation}
       {\cal D}(H)=\mathcal{H}_{\rm b} \oplus {\cal D}(H_{\rm c}) 
\end{equation}
and denote the restriction of $H$ to ${\cal D}(H_{\rm c})$ by $H_{\rm c}$, 
then the operator $\widetilde{U}_{\rm c}$ of Eq.~(\ref{rhoU}) provides a 
$\rho$-diagonalization of $H_{\rm c}$.  If we 
seek a $\delta$-normalization~\cite{DIS}, we just have to define the following
eigenfunction:
\begin{equation}
      \phi (r;E):=\sqrt{\rho (E)} \, \chi (r;E) \, ,
      \label{dnes}
\end{equation}
which is the eigensolution of the differential operator $h$ that 
is $\delta$-normalized, and the following unitary operator:
\begin{eqnarray}
      U_{\rm c}:\mathcal{H}_{\rm c} 
      &\longmapsto & \widehat{\mathcal{H}}_{\rm c} \nonumber \\
       f_{\rm c} &\longmapsto & \widehat{f}_{\rm c}(E)=
        (U_{\rm c}f_{\rm c})(E)=\int_0^{\infty}dr \, 
         f_{\rm c}(r) \overline{\phi (r;E)} \, ,
\end{eqnarray}
where $\widehat{\mathcal{H}}_{\rm c}=L^2([0,\infty ),dE)$.  We note that the 
normalization of the bound states (\ref{normaiegin}) is different to the 
$\delta$-normalization of (\ref{dnes}).

The inverses of the operators $U_{\rm b}$ and $U_{\rm c}$ are provided by the 
following theorem (cf.~Theorem XIII.5.14 of Ref.~\cite{DUNFORD}):

\vskip0.5cm

{\bf Theorem~4} (Weyl-Kodaira)\quad  Let $H$, $\Lambda$, 
$\{ \rho _{ij} \}$, etc., be as in Theorem~2.  Let $E_0$ and $E_1$ be the end 
points of $\Lambda$.  Then the inverse of the isometric isomorphism $U$ of 
${\sf E}(\Lambda )L^2([0,\infty ),dr)$ onto $L^2(\Lambda ,\{ \rho _{ij}\})$ is 
given by the formula
\begin{equation}
       (U^{-1}F)(r)=\lim_{\mu _0 \to E_0}\lim_{\mu _1 \to E_1}
       \int_{\mu _0}^{\mu _1} \left( \sum_{i,j=1}^{2}
             F_i(E)\sigma _j(r;E)\rho _{ij}(dE) \right) \, ,
\end{equation}
where $F=[F_1,F_2]\in L^2(\Lambda ,\{ \rho _{ij}\})$, the limit existing
in the topology of $L^2([0, \infty ),dr)$.

\vskip0.5cm

According to Theorem~4, the inverse of $U_{\rm b}$ is given by 
\begin{equation}
      (U_{\rm b}^{-1}\widehat{f}_{\rm b})(r)=
      \sum_{n=1}^N (\phi _n,f_{\rm b}) \phi _n(r) \, ,
\end{equation}
where 
$\widehat{f}_{\rm b} =\{ (\phi _1,f_{\rm b}), \ldots , (\phi _N,f_{\rm b}) \}$
is an $N$-tuple of complex numbers.  The operator $U_{\rm b}^{-1}$ transforms
from $\widehat{\mathcal{H}}_{\rm b}$ onto $\mathcal{H}_{\rm b}$.

According to Theorem~4, the inverse of $U_{\rm c}$ is given by 
\begin{equation}
      f_{\rm c}(r)=(U_{\rm c}^{-1}\widehat{f}_{\rm c})(r)= 
        \int_0^{\infty}dE\, \widehat{f}_{\rm c}(E)\phi (r;E) \, ,
       \quad \widehat{f}_{\rm c}(E)\in L^2([0,\infty ),dE) \, .
     \label{invdiagonaliza}
\end{equation}
The operator $U_{\rm c}^{-1}$ transforms
from $\widehat{\mathcal{H}}_{\rm c}$ onto $\mathcal{H}_{\rm c}$.

Using the operators $U_{\rm b}$ and $U_{\rm c}$ we can construct an operator
$U$ defined as 
\begin{eqnarray}
      U:{\cal H} &\longmapsto & 
      \widehat{\cal H}=
      \oplus _{n=1}^N \widehat{\mathcal H}(E_n) \oplus 
       \int_{0}^{\infty} \widehat{\cal H}(E)dE 
             \nonumber \\
       f &\longmapsto & Uf:= [U_{\rm b}f,U_{\rm c}f]= 
       \left[  \widehat{f}_{\rm b},\{ \widehat{f}_{\rm c}(E) \} \right] \, ,
       \label{dirintdec}
\end{eqnarray}
where ${\cal H}$ is realized by 
$\mathcal{H}_{\rm b} \oplus \mathcal{H}_{\rm c}$, and $\widehat{\cal H}$ is 
realized by 
$\widehat{\mathcal{H}}_{\rm b} \oplus \widehat{\mathcal{H}}_{\rm c}$.  The
Hilbert space $\widehat{\mathcal H}(E_n)$, which is associated to each
energy $E_n$ in the discrete spectrum, and the Hilbert space 
$\widehat{\cal H}(E)$, which is associated to each energy $E$ in the 
continuous spectrum, are realized by the Hilbert space of complex numbers 
$\mathbb C$.  On $\widehat{\cal H}$, the operator $H$ acts as the 
multiplication operator,
\begin{equation}
       Hf \longmapsto UHf\equiv \sum_{n=1}^N E_n(\phi _n,f_{\rm b}) + 
          \{ E\widehat{f}_{\rm c}(E) \} 
         \, , \quad f\in {\cal D}(H) \, . 
\end{equation}
The scalar product on $\widehat{\cal H}$ can be written as
\begin{equation}
       \left( \widehat{f},\widehat{g} \right) _{\widehat{\cal H}}=
        \sum_{n=1}^N (f_{\rm b},\phi _n)(\phi _n,g_{\rm b})+
       \int_0^{\infty}dE\,  
       \overline{\widehat{f}_{\rm c}(E)} \, \widehat{g}_{\rm c}(E) \, ,
\end{equation}
where 
$\widehat{f}_{\rm b}=\{(\phi _1,f_{\rm b}), \ldots , (\phi _N,f_{\rm b})\}$
and 
$\widehat{g}_{\rm b}=\{(\phi _1,g_{\rm b}), \ldots , (\phi _N,g_{\rm b})\}$.  

Therefore, $U$ induces a direct integral decomposition of the Hilbert space 
associated to the Hamiltonian $H$ (see~\cite{VON2,GELFAND}).  In particular, 
we can write the eigenfunction expansions of any element of the Hilbert spaces 
$L^2([0,\infty ),dr )$ and $L^2([0,\infty ),dE)$ in terms of the 
eigensolutions $\phi _n(r)$ and $\phi (r;E)$ of $h$,
\begin{equation}
       f(r)=f_{\rm b}(r)+f_{\rm c}(r)=
            \sum_{n=1}^N(\phi _n,f)\phi_n(r) +
            \int_0^{\infty}dE\, \widehat{f}(E) \phi (r;E) \, , 
       \label{eigenfunex1}
\end{equation}
\begin{equation}
       \widehat{f}\equiv \widehat{f}_{\rm b}+\widehat{f}_{\rm c}(E)=
         \{ (\phi _n,f)\}_{n=1}^N +
       \int_0^{\infty}dr\, f(r) \overline{\phi (r;E)} \, .
       \label{eigenfunex2}
\end{equation}

\def\thesection{\arabic{section}}
\section{Rigged Hilbert Space}
\def\thesection{\arabic{section}}
\setcounter{equation}{0}
\label{sec:constRHS}

As explained in the Introduction, the Hilbert space framework (in particular,
the direct integral decomposition) is not sufficient for the purposes of 
Quantum Mechanics: an extension of the Hilbert space to the RHS is needed.  To
construct the RHS, we shall consider the discrete and the continuous spectra 
separately. 

\subsection{Construction of the Rigged Hilbert Space}
\label{sec:RotD}

First, we construct the RHS associated to the discrete spectrum.  In this case,
the Hilbert space ${\cal H}_{\rm b}$ is finite dimensional.  Therefore, all 
the difficulties exposed in the Introduction do not arise.  The ``Gelfand 
triplet'' that corresponds to the discrete spectrum consists of three copies 
of the Hilbert space ${\cal H}_{\rm b}$,
\begin{equation}
  \mathbf \Phi _{\rm b}=\mathcal{H}_{\rm b}=\mathbf \Phi _{\rm b}^{\times}\, .
  \label{RHSdisc}
\end{equation}
The eigenket $|E_n)$ associated to any energy $E_n$ in the discrete spectrum
is defined by
\begin{eqnarray}
       |E_n) :\mathbf \Phi _{\rm b}& \longmapsto & {\mathbb C} \nonumber \\
       \varphi _{\rm b}& \longmapsto & (\varphi _{\rm b}|E_n) 
       :=(\varphi _{\rm b},\phi _n)= 
       \int_0^{\infty}dr \, \overline{\varphi _{\rm b}(r)}\phi _n(r) \, .
       \label{bneedefinitionket}
\end{eqnarray}
Clearly, the function $|E_n)$ is an antilinear functional over 
$\mathbf \Phi _{\rm b}$, i.e., 
$|E_n)\in \mathbf \Phi _{\rm b}^{\times}$.  Besides being an eigenvector
of the Hamiltonian in the usual sense, $|E_n)$ is also an eigenvector of
$H_{\rm b}$ in the RHS sense of Eq.~(\ref{idef}),
\begin{equation}
       (\varphi _{\rm b}|H_{\rm b}^{\times}|E_n)=E_n(\varphi _{\rm b}|E_n) 
       \, , 
       \quad \forall \varphi _{\rm b} \in \mathbf \Phi _{\rm b} \, .
\end{equation}
Although it is not necessary to use the RHS formalism when dealing with the 
discrete spectrum, we have constructed the ``RHS'' (\ref{RHSdisc}) and the
eigenket (\ref{bneedefinitionket}) in order to make a parallel to the 
continuous case.

Now, we construct the RHS associated to the continuous 
spectrum.  The first step is to make all the powers of the Hamiltonian 
well-defined.  In order to do so, we construct the maximal invariant subspace 
${\cal D}_{\rm c}$ of the operator $H_{\rm c}$,
\begin{equation}
      {\cal D}_{\rm c}:= \bigcap _{n=0}^{\infty}{\cal D}(H_{\rm c}^n) \, .
      \label{misus}
\end{equation}
The space ${\cal D}_{\rm c}$ is the largest subspace of ${\cal D}(H_{\rm c})$ 
that remains stable under the action of the Hamiltonian $H_{\rm c}$ and of
its powers.  Recalling that $h$ denotes the formal differential operator
(\ref{doh}), it is easy to check that
\begin{eqnarray}
      {\cal D}_{\rm c}=\{ \varphi _{\rm c}\in {\cal H}_{\rm c} \, | && \ 
       h^n\varphi _{\rm c}(r)\in {\cal H}_{\rm c},\
       h^n\varphi _{\rm c}(0)=0, \ 
      \varphi _{\rm c}^{(n)}(a)=\varphi _{\rm c}^{(n)}(b)=0, \
       n=0,1,2,\ldots ; \nonumber \\
      && \varphi _{\rm c}(r) \in C^{\infty}([0,\infty)) \} 
          \, .
      \label{mainisexi}
\end{eqnarray}

The second step is to find the subspace 
$\mathbf \Phi _{\rm c}\subset {\cal D}_{\rm c}$ on which
the eigenkets $|E\rangle$ of Eq.~(\ref{Iceigenkets}) act as 
antilinear functionals.  Those eigenkets are to be defined as integral 
operators whose kernel is the eigenfunction $\phi (r;E)$,
\begin{eqnarray}
       |E\rangle :\mathbf \Phi _{\rm c}& \longmapsto & {\mathbb C} \nonumber \\
       \varphi _{\rm c}& \longmapsto & \langle \varphi _{\rm c}|E\rangle := 
       \int_0^{\infty}dr\, \overline{\varphi _{\rm c}(r)}\phi (r;E) 
       =\overline{(U_{\rm c}\varphi _{\rm c})(E)} \, .
       \label{definitionket}
\end{eqnarray}
A necessary condition for $|E\rangle$ to be
well-defined is that $\mathbf \Phi _{\rm c}$ be included in 
${\cal D}_{\rm c}$.  This condition, however, is not sufficient to obtain
a continuous functional.  In order to make the eigenfunctional $|E\rangle$ 
continuous, we have to impose further restrictions on the behavior at infinity
of the elements of ${\cal D}_{\rm c}$,
\begin{equation}
      \int_0^{\infty}dr \, 
        \left| (r+1)^n(h+1)^m\varphi _{\rm c}(r)\right| ^2<\infty \, , 
        \quad n,m=0,1,2,\ldots
       \label{condcodogis}
\end{equation}
The imposition of (\ref{condcodogis}) upon the space 
${\cal D}_{\rm c}$ yields
\begin{equation}
       {\mathbf \Phi}_{\rm c}=\{ \varphi _{\rm c}\in {\cal D}_{\rm c} \, | \ 
       \int_0^{\infty}dr \, 
       \left| (r+1)^n(h+1)^m\varphi _{\rm c}(r)\right| ^2<\infty, 
        \quad n,m=0,1,2,\ldots \} \, .
\end{equation}    
On $\mathbf \Phi _{\rm c}$, we define the family of norms
\begin{equation}
      \| \varphi _{\rm c}\| _{n,m} := 
 \sqrt{\int_0^{\infty}dr \, \left| (r+1)^n(h+1)^m\varphi _{\rm c}(r)\right| ^2}
    \, , \quad n,m=0,1,2,\ldots 
      \label{nmnorms}
\end{equation}
These norms can be used to define a countably normed topology 
$\tau _{\mathbf \Phi _{\rm c}}$ on $\mathbf \Phi _{\rm c}$ 
(see~\cite{GELFAND}),  
\begin{equation}
      \varphi _{{\rm c},\alpha}\, 
       \mapupdown{\tau_{\mathbf \Phi _{\rm c}}}{\alpha \to \infty}
      \, \varphi _{\rm c}\quad {\rm iff} \quad  
      \| \varphi _{{\rm c},\alpha }-\varphi _{\rm c}\| _{n,m} 
      \, \mapupdown{}{\alpha \to \infty}\, 0 \, , \quad n,m=0,1,2, \ldots 
\end{equation}       

Once we have constructed the space $\mathbf \Phi _{\rm c}$, we can construct 
its topological dual $\mathbf \Phi _{\rm c}^{\times}$ as the space of 
$\tau _{\mathbf \Phi _{\rm c}}$-continuous antilinear functionals on 
$\mathbf \Phi _{\rm c}$ (see~\cite{GELFAND}) and therewith the RHS 
corresponding to the continuous part of the spectrum,
\begin{equation}
       \mathbf \Phi _{\rm c}\subset {\cal H} _{\rm c}\subset 
       \mathbf \Phi _{\rm c}^{\times}
        \, .
       \label{RHSCONT}
\end{equation}

In order to show that the RHS (\ref{RHSCONT}) is what we are looking for,
we need to prove the following proposition (the proof can be found
in Appendix~\ref{sec:appendixB}):

\vskip0.5cm

{\bf Proposition~1} \quad The triplet of spaces (\ref{RHSCONT}) is a 
Rigged Hilbert Space, and it satisfies all the requirements demanded 
in the Introduction.  More specifically,

\vskip0.3cm
  
(i) The quantities (\ref{nmnorms}) fulfill the conditions to be a norm. 

\vskip0.3cm

(ii) The space $\mathbf \Phi _{\rm c}$ is stable under the action of 
$H_{\rm c}$, and $H_{\rm c}$ is $\tau _{\mathbf \Phi _{\rm c}}$-continuous.

\vskip0.3cm

(iii) The ket $|E\rangle$ of Eq.~(\ref{definitionket}) is a 
well-defined antilinear functional on $\mathbf \Phi _{\rm c}$, i.e., 
$|E\rangle$ belongs to $\mathbf \Phi _{\rm c}^{\times}$.  

\vskip0.3cm

(iv) The ket $|E\rangle$ is a generalized eigenvector of $H_{\rm c}$,
\begin{equation}
       H_{\rm c}^{\times}|E\rangle=E|E\rangle \, ,
\end{equation}
i.e.,
\begin{equation}
       \langle \varphi _{\rm c}|H_{\rm c}^{\times}|E\rangle=
        \langle H_{\rm c}^{\dagger}\varphi _{\rm c} |E\rangle = 
      E\langle \varphi _{\rm c}|E\rangle \, ,
       \quad \forall \varphi _{\rm c}\in \mathbf \Phi _{\rm c} \, .
       \label{nafegenphis}
\end{equation}        

\vskip0.5cm

(This proposition is a generalization to infinite dimensional
spaces of the results that hold in finite dimensional spaces.)

We now combine the RHSs (\ref{RHSdisc}) and (\ref{RHSCONT})
corresponding to the discrete and continuous spectrum into the RHS of the 
square well-barrier potential,
\begin{equation}
       \mathbf \Phi \subset {\cal H} \subset \mathbf \Phi ^{\times} \, ,
       \label{ABRHSCONT}
\end{equation}
where
\begin{equation}
       \mathbf \Phi :=\mathbf \Phi _{\rm b}\oplus \mathbf \Phi _{\rm c} \, ; 
       \qquad {\cal H}:={\cal H}_{\rm b}\oplus {\cal H}_{\rm c} \, ; \qquad
  \mathbf \Phi ^{\times} :=\mathbf \Phi _{\rm b}^{\times}
        \oplus \mathbf \Phi _{\rm c}^{\times} \, .
\end{equation}
On the space $\mathbf \Phi$, all the expectation values of the 
Hamiltonian and all the algebraic operations involving $H$ are 
well-defined, and the eigenvalue equations (\ref{Ieigenkets}) 
hold.  The kets $|E_n)$ and $|E\rangle$ satisfy the normalization 
(\ref{Ideltanorintro}).  Since the spaces $\mathbf \Phi _{\rm b}$ and 
$\mathbf \Phi _{\rm c}$ are orthogonal to each other, we also have that
\begin{equation}
       (\varphi|E_n)=(\varphi _{\rm b}|E_n) \, ;  \quad 
      \langle \varphi|E\rangle =\langle \varphi _{\rm c}|E\rangle \, .
\end{equation}
As we shall see in the next section, the kets $|E_n)$ and $|E\rangle$ form a 
complete basis system.

Before finishing this section, we would like to remark that $\phi (r;E)$ is
\emph{not} the same object as the ket $|E\rangle$ of 
Eq.~(\ref{definitionket}).  The function $\phi (r;E)$ is an eigenfunction of 
the formal differential operator $h$, whereas the ket $|E\rangle$ is a 
generalized eigenvector of the Hamiltonian.  The eigenket $|E\rangle$ is 
defined by the eigenfunction $\phi (r;E)$ \emph{and} by the space of test 
functions $\mathbf \Phi _{\rm c}$ on which it acts as an antilinear 
functional.

\subsection{Dirac Basis Vector Expansion}

In the Introduction, we mentioned that in Quantum Mechanics it is
assumed that the eigenkets of the Hamiltonian form a complete basis system
that can be used to expand any physical wave function $\varphi$ as in
Eq.~(\ref{IDiracbsis}).  This expansion is derived in the present 
section.  That derivation consists of the restriction of the eigenfunction 
expansion (\ref{eigenfunex1}) to the space $\mathbf \Phi$.  

If we denote $\langle r|\varphi \rangle \equiv \varphi (r)$, 
$\langle r|E_n)\equiv \phi _n(r)$,  and $\langle r|E\rangle \equiv \phi (r;E)$,
and if we define the action of the bra $\langle E|$ on the wave function
$\varphi$ by 
$\langle E| \varphi \rangle := \overline{\langle \varphi|E\rangle}$, then
Eq.~(\ref{eigenfunex1}) can be written as
\begin{equation}
      \langle r|\varphi \rangle =\sum_{n=0}^{\infty}
              \langle r|E_n) (E_n|\varphi )+   
         \int_0^{\infty}dE \,
          \langle r|E \rangle \langle E|\varphi \rangle \, , \quad
       \varphi \in \mathbf \Phi \, .
       \label{inveqDva}
\end{equation}
This is the Dirac basis vector expansion of the square well-barrier 
potential.  Although the 
eigenfunction expansion (\ref{eigenfunex1}) is valid for every element of the 
Hilbert space, Dirac basis vector expansion (\ref{inveqDva}) is only 
valid for functions $\varphi \in  \mathbf \Phi$, because only those 
functions fulfill both
\begin{equation}
      \overline{\widehat{\varphi}_{\rm c}(E)}=\langle \varphi |E\rangle
\end{equation}
and
\begin{equation}
       \langle \varphi |H^{\times}|E\rangle = 
        \langle H^{\dagger}\varphi |E\rangle =
       E \langle \varphi|E\rangle \, .
\end{equation}

For the sake of completeness, we include the Nuclear Spectral 
Theorem~\cite{GELFAND}, which is usually referred to as the mathematical
justification of the heuristic Dirac basis vector 
expansion~\cite{ANTOINE,B60,ROBERTS,GELFAND,GALINDO,MELSH}.  Instead of using 
the general proof of Ref.~\cite{GELFAND}, we prove that theorem using the 
machinery of the Sturm-Liouville theory (our proof can be found in 
Appendix~\ref{sec:appendixB}).  

\vskip0.5cm

{\bf Proposition~2} (Nuclear Spectral Theorem) \quad Let 
\begin{equation}
     {\mathbf \Phi} \subset L^2([0,\infty ),dr)\subset {\mathbf \Phi}^{\times}
\end{equation}
be the RHS of the square well-barrier Hamiltonian $H$ such that $\mathbf \Phi$ 
remains invariant under $H$ and $H$ is a $\tau _{\mathbf \Phi}$-continuous 
operator on $\mathbf \Phi$.  Then, for each energy in 
the spectrum of $H$ there is a generalized eigenvector such that
\begin{mathletters}
\begin{eqnarray}
       &H|E_n) =E_n|E_n) \, , \quad &E_n \in \{ E_1, \ldots ,E_N \} \, , \\
       &H^{\times}|E\rangle =E|E\rangle \, , \quad &E\in [0,\infty ) \, , 
\end{eqnarray}
\end{mathletters}
and such that
\begin{equation}
      (\varphi ,\psi )=\sum_{n=1}^N(\varphi |E_n)(E_n|\psi )+ 
       \int_0^{\infty}dE\, 
      \langle \varphi |E\rangle \langle E|\psi \rangle \, , \quad 
       \forall \varphi ,\psi \in \mathbf \Phi \, ,
       \label{GMT1}
\end{equation}
and
\begin{equation}
      (\varphi ,H^m \psi )=\sum_{n=1}^NE_n^m(\varphi |E_n)(E_n|\psi )+ 
     \int_0^{\infty} dE \,
      E^m \langle \varphi |E\rangle \langle E|\psi \rangle \, , \quad 
       \forall \varphi ,\psi \in {\mathbf \Phi} \, , m=1,2,\ldots
        \label{GMT2}
\end{equation}

\vskip0.5cm

Thus this theorem allows us to write the scalar product (\ref{GMT1}) of any 
two functions $\varphi ,\psi$ of $\mathbf \Phi$ and the matrix elements
(\ref{GMT2}) in terms of the action of the kets $|E_n)$ and $|E\rangle$ on 
$\varphi ,\psi$.

\subsection{Energy and Momentum Representations of the Rigged Hilbert Space}

In this section, we construct the energy representation of the RHS 
(\ref{ABRHSCONT}) by applying to it the unitary operator $U$ of 
Eq.~(\ref{dirintdec}).  As done throughout this paper, we study the discrete
and the continuous case separately. 

In the discrete case (\ref{RHSdisc}), we have already seen that the energy
representation of ${\cal H}_{\rm b}$ is ${\mathbb C}^N$.  Hence the energy
representation of the RHS (\ref{RHSdisc}) consists of three copies of the 
space ${\mathbb C}^N$.  The image of the eigenket $|E_n)$ under 
$U_{\rm b}^{\times}$ is the $N$-tuple that has zeros everywhere but at the 
$n$-th position, $\{ \delta _{in} \}_{i=1}^N$.  In the energy representation, the Hamiltonian
$H_{\rm b}$ acts as the diagonal matrix (\ref{diagonaHb}).

In the continuous case (\ref{RHSCONT}), we have already shown that in the 
energy representation the Hamiltonian $H_{\rm c}$ acts as the multiplication 
operator.  The energy representation of the space $\mathbf \Phi _{\rm c}$ is 
defined as  
\begin{equation}
      \widehat{\mathbf \Phi}_{\rm c}:= U_{\rm c}\mathbf \Phi _{\rm c}\, .
\end{equation}
The space $\widehat{\mathbf \Phi}_{\rm c}$ is a linear subspace of 
$L^2([0,\infty ),dE)$.  In order to endow $\widehat{\mathbf \Phi}_{\rm c}$ with
a topology $\tau _{\widehat{\mathbf \Phi}_{\rm c}}$, we carry the topology on 
$\mathbf \Phi _{\rm c}$ into $\widehat{\mathbf \Phi}_{\rm c}$,
\begin{equation}
      \tau _{\widehat{\mathbf \Phi}_{\rm c}}:=
      U_{\rm c}\tau _{\mathbf \Phi _{\rm c}} \, .
\end{equation}
With this topology, the space $\widehat{\mathbf \Phi}_{\rm c}$ is a linear 
topological space.  If we denote the dual space of 
$\widehat{\mathbf \Phi}_{\rm c}$ by $\widehat{\mathbf \Phi}_{\rm c}^{\times}$,
then we have
\begin{equation}
      U_{\rm c}^{\times}{\mathbf \Phi}_{\rm c}^{\times}=
      (U_{\rm c}{\mathbf \Phi}_{\rm c})^{\times}= 
       \widehat{\mathbf \Phi}_{\rm c}^{\times} \, .
\end{equation}
From the following chain of equalities:
\begin{eqnarray}
       \langle \widehat{\varphi }_{\rm c}|U_{\rm c}^{\times}|E\rangle &=&
       \langle U_{\rm c}^{-1}\widehat{\varphi}_{\rm c}|E\rangle  \nonumber \\
     &=& \int_0^{\infty}dr\, \overline{\varphi _{\rm c}(r)}\phi (r;E)
          \nonumber \\
       &=& \overline{\widehat{\varphi}_{\rm c}(E)} \, ,
\end{eqnarray}
it follows that the energy representation of the Dirac ket $|E\rangle$, 
that we denote by $|\widehat{E}\rangle$, is the antilinear Schwartz delta 
functional, i.e., $|\widehat{E}\rangle$ is the antilinear functional that 
associates to each function $\widehat{\varphi}_{\rm c}$ the complex conjugate 
of its value at the point $E$.

In order to summarize all these results, it is very helpful to show the 
different realizations of the RHS through the following diagram:
\begin{equation}
      \begin{array}{cccccccccc}
      H; & \varphi & \ & \mathbf \Phi_{\rm b}\oplus \mathbf \Phi_{\rm c} &
      \subset & {\cal H}_{\rm b}\oplus {\cal H}_{\rm c}  &
 \subset &\mathbf \Phi_{\rm b}^{\times}\oplus \mathbf \Phi_{\rm c}^{\times} &
   \  & |E_n)\oplus |E\rangle \nonumber \\  
       & & \ & U_{\rm b} \downarrow \oplus \downarrow U_{\rm c}  &  &
       U_{\rm b}\downarrow \oplus \downarrow  U_{\rm c}   &
       &U_{\rm b}^{\times} \downarrow \oplus \downarrow U_{\rm c}^{\times} & 
        \ &   \nonumber \\   
      \widehat{H}; & \widehat{\varphi} & \ &
      \widehat{\mathbf \Phi}_{\rm b}\oplus \widehat{\mathbf \Phi}_{\rm c}& 
       \subset & 
      \widehat{{\cal H}}_{\rm b}\oplus \widehat{{\cal H}}_{\rm c} & \subset & 
   \widehat{\mathbf \Phi}_{\rm b}^{\times}\oplus 
    \widehat{\mathbf \Phi}_{\rm c}^{\times} & 
      \ & |\widehat{E}_n)\oplus |\widehat{E}\rangle  \\ 
      \end{array}
      \label{diagramsavp}
\end{equation}
On the top line of the diagram (\ref{diagramsavp}), we have the RHS, the 
Hamiltonian, the wave functions and the Dirac kets in the position 
representation.  On the bottom line, we have their energy representation 
counterparts.

Since every function we have computed (eigenfunctions of the 
Schr\"odinger differential operator, Green function, etc.) depends on the
square root of the energy rather than on the energy itself, the momentum $k$
is a more convenient variable, especially for the continuous case.  The 
momentum $k$ is defined by
\begin{equation}
        k:=\sqrt{\frac{2m}{\hbar ^2} \, E} \, . 
       \label{momenuks}
\end{equation}
In terms of $k$, the $\delta$-normalized eigensolution of the differential
operator (\ref{doh}) reads
\begin{equation}
      \langle r|k\rangle =
     \left[2\pi {\cal J}_3(k){\cal J}_4(k) \right]^{-1/2} \chi (r;k) 
      \, . \label{continuoseign}
\end{equation}
This eigensolution can be used to obtain the unitary operator $V_{\rm c}$ 
that transforms from the position into the momentum representation,
\begin{equation}  
     \widehat{f}_{\rm c}(k)=(V_{\rm c}f_{\rm c})(k)=
      \int_0^{\infty}dr\, f_{\rm c}(r) \overline{\langle r|k\rangle}\, ,
       \quad f_{\rm c}\in {\cal H}_{\rm c} \, . 
    \label{Vcontinuoseign}
\end{equation}
(The operator $V_{\rm b}$ corresponding to the discrete case is
$U_{\rm b}$.)  On the $k$-representation, the (continuous part of the) 
Hamiltonian acts as multiplication by $k^2$.  To each $k\in [0,\infty )$, there
corresponds an eigenket $|k\rangle$ that acts on $\mathbf \Phi _{\rm c}$ as 
the following integral operator:  
\begin{equation}
      \langle \varphi _{\rm c}|k\rangle :=\int_0^{\infty}dr\, 
       \langle \varphi _{\rm c}|r\rangle \langle r|k\rangle=
       \overline{(V_{\rm c}\varphi _{\rm c})(k)}\, , \quad
        \varphi _{\rm c} \in \mathbf \Phi _{\rm c}  \, . 
\end{equation}
This eigenket is a generalized eigenvector of the Hamiltonian corresponding
to the eigenvalue $k^2$.  In this way, we can construct the 
$k$-representation of all the objects of the diagram (\ref{diagramsavp}).

If we express our functions in terms of $k$, there is no need to distinguish 
different regions in the $E$-plane, because in the $k$-plane the resolvent and
the $S$-matrix do not have a cut.  There is no need to work with a 
particular branch of the square root either, because Eq.~(\ref{momenuks}) 
provides a Riemann surface in a natural way.  For instance, we can define the 
Jost function
\begin{equation}
      {\cal J}(k):={\cal J}_+(k) \, , \quad k\in {\mathbb C} \, ,
      \label{junfcion}
\end{equation}
the eigensolution
\begin{equation}
      f(r;k):=\Theta _+(r;k) \, , \quad k\in {\mathbb C} \, , 
\end{equation}
and the Green function
\begin{equation}
      G(r,r';k):=-\frac{2m/\hbar ^2}{\sqrt{2m/\hbar ^2 \, E}}
         \frac{\chi (r_<;k)f(r_>;k)}{{\cal J}(k)} \, , 
        \quad k\in {\mathbb C} \, ,  
      \label{grenkfunc}
\end{equation}
where $r_<$, $r_>$ refer to the smaller and to the bigger of $r$ and $r'$,
respectively.  Then, as $k$ approaches the real positive
$k$ axis, or as $E$ approaches the right-hand cut from above, 
${\cal J}(k)$, $f(r;k)$ and $G(r,r';k)$ become 
${\cal J}_+(k)$, $\Theta _+(r;k)$ and $G^+(r,r';k)$, respectively; as $k$ 
approaches the negative real axis (from above), or $E$ the right-hand
cut from below, ${\cal J}(k)$, $f(r;k)$ and $G(r,r';k)$ become  
${\cal J}_-(k)$, $\Theta _-(r;k)$ and $G^-(r,r';k)$, 
respectively~\cite{NEWTON}.  Also, the values that the functions of 
Eqs.~(\ref{junfcion})-(\ref{grenkfunc}) take on the second sheet of the 
Riemann surface are already specified (the Sturm-Liouville theory deals only 
with energies in the first sheet).  This is particularly useful when studying 
resonance energies, which lie on the second 
sheet~\cite{MONDRAGON1,MONDRAGON2,MONDRAGON3}.

\def\thesection{\arabic{section}}
\section{Conclusions}
\def\thesection{\arabic{section}}
\setcounter{equation}{0}
\label{sec:conclusions}

We have shown that the solutions of the Schr\"odinger equation corresponding
to the square well-barrier potential fall in the Rigged Hilbert Space
\begin{equation}
      \mathbf \Phi _{\rm b} \oplus \mathbf \Phi _{\rm c} \subset 
      {\cal H}_{\rm b} 
     \oplus {\cal H}_{\rm c}
      \subset \mathbf \Phi _{\rm b}^{\times} \oplus 
      \mathbf \Phi _{\rm c}^{\times} \, .  
       \label{rihscon}
\end{equation}
The spectrum of the Hamiltonian has a discrete part
$\{ E_1, \ldots ,E_N \}$ and a continuous part $[0,\infty)$.  For each
energy in the spectrum of $H$, we have constructed an eigenket that is
an eigenvector of $H$.  If the energy $E_n$ belongs to the discrete spectrum,
its corresponding eigenket $|E_n)$, which is given by 
Eq.~(\ref{bneedefinitionket}), is an element of 
$\mathbf \Phi _{\rm b}^{\times}$.  If the energy $E$ 
belongs to the continuous spectrum, its corresponding eigenket $|E\rangle$,
which is given by Eq~(\ref{definitionket}), is an element of
$\mathbf \Phi _{\rm c}^{\times}$.  Thus 
the RHS treats the discrete and the continuous spectra independently and on 
the same footing.  Each element $\varphi$ of 
$\mathbf \Phi _{\rm b} \oplus \mathbf \Phi _{\rm c}$ can be expanded by the
eigenkets of the Hamiltonian as in Eq.~(\ref{inveqDva}).  The elements of 
$\mathbf \Phi _{\rm b} \oplus \mathbf \Phi _{\rm c}$ are the only square
normalizable functions for which such an expansion is possible.  The 
expectation values and the uncertainties of the Hamiltonian are well defined 
quantities in each element $\varphi$ of 
$\mathbf \Phi _{\rm b} \oplus \mathbf \Phi _{\rm c}$.

Therefore, the RHS (\ref{rihscon}) contains all the physically meaningful
solutions of the Schr\"odinger equation---the physical wave functions are
included in $\mathbf \Phi _{\rm b} \oplus \mathbf \Phi _{\rm c}$, whereas the
monoenergetic solutions of the time independent Schr\"odinger equation are
included in 
$\mathbf \Phi _{\rm b}^{\times} \oplus \mathbf \Phi _{\rm c}^{\times}$.

Although illustrated within the example of the square well-barrier potential, 
these conclusions hold in general: the solutions of the Schr\"odinger equation
fall in a RHS rather than just in a Hilbert space.

\section*{Acknowledgments}

The author thanks C.~Koeninger, N.~L.~Harshman, and Profs.~A.~Bohm, M.~Gadella
and A.~Galindo for invaluable comments.  Financial 
support from the U.E.~TMR Contract number ERBFMRX-CT96-0087 ``The 
Physics of Quantum Information'' is gratefully acknowledged.

\appendix
\def\thesection{\Alph{section}}
\section{List of Auxiliary Functions}
\setcounter{equation}{0}
\label{sec:appendix}

If we define 
\begin{equation}
       \widetilde{k} :=\sqrt{-\frac{2m}{\hbar ^2}\, E} \, ; \quad 
       \widetilde{Q}_1:=\sqrt{-\frac{2m}{\hbar ^2}\, (E+V_1)} \, ; \quad 
       \widetilde{Q}_2:=\sqrt{-\frac{2m}{\hbar ^2}\, (E-V_2)} \, .
\end{equation}
then the coefficients in Eq.~(\ref{tildechifunction}) are given by
\begin{mathletters}
\begin{eqnarray}
      &&\widetilde{{\cal J}}_1(E)\equiv \widetilde{{\cal J}}_1(\widetilde{k})=
       \frac{i}{4}e^{-\widetilde{Q}_2a}
             \left[ 
     \left(1+\frac{\widetilde{Q}_1}{\widetilde{Q}_2}\right) 
     e^{\widetilde{Q}_1a} -
     \left(1-\frac{\widetilde{Q}_1}{\widetilde{Q}_2} \right)
     e^{-\widetilde{Q}_1a} \right] , \hskip1cm \\
     && \widetilde{{\cal J}}_2(E)\equiv\widetilde{{\cal J}}_2(\widetilde{k})=
       \frac{i}{4}e^{\widetilde{Q}_2a}
             \left[ 
     \left(1-\frac{\widetilde{Q}_1}{\widetilde{Q}_2}\right) 
     e^{\widetilde{Q}_1a} -
     \left(1+\frac{\widetilde{Q}_1}{\widetilde{Q}_2} \right)
     e^{-\widetilde{Q}_1a} \right] , \hskip1cm \\
     && \widetilde{{\cal J}}_3(E)\equiv \widetilde{{\cal J}}_3(\widetilde{k})=
             \frac{1}{2}e^{-\widetilde{k}b}
                   \left[
      \left(1+\frac{\widetilde{Q}_2}{\widetilde{k}}\right) 
      e^{\widetilde{Q}_2b}\widetilde{{\cal J}}_1(\widetilde{k})+
     \left(1-\frac{\widetilde{Q}_2}{\widetilde{k}} \right) 
      e^{-\widetilde{Q}_2b}\widetilde{{\cal J}}_2(\widetilde{k})
     \right]  , \hskip1cm \\ 
     &&\widetilde{{\cal J}}_4(E)\equiv \widetilde{{\cal J}}_4(\widetilde{k})=
               \frac{1}{2}e^{\widetilde{k}b}
                   \left[
     \left(1-\frac{\widetilde{Q}_2}{\widetilde{k}} \right) 
     e^{\widetilde{Q}_2b}\widetilde{{\cal J}}_1(\widetilde{k})+
     \left(1+\frac{\widetilde{Q}_2}{\widetilde{k}} \right) 
      e^{-\widetilde{Q}_2b}\widetilde{{\cal J}}_2(\widetilde{k})
     \right]  ,  \hskip1cm
\end{eqnarray}
     \label{tildejfunc}
\end{mathletters}
and the coefficients in Eq.~(\ref{tildethetfunc}) by
\begin{mathletters}
\begin{eqnarray}
      && \widetilde{{\cal A}}_3(E)\equiv \widetilde{{\cal A}}_3(\widetilde{k})=
      \frac{1}{2}e^{-\widetilde{Q}_2b} 
     \left(1-\frac{\widetilde{k}}{\widetilde{Q}_2}\right) 
     e^{-\widetilde{k}b} , \hskip1.5cm \\
     && \widetilde{{\cal A}}_4(E)\equiv \widetilde{{\cal A}}_4(\widetilde{k})=
       \frac{1}{2}e^{\widetilde{Q}_2b}
       \left(1+\frac{\widetilde{k}}
            {\widetilde{Q}_2} \right)
       e^{-\widetilde{k}b} , \hskip1.5cm \\
    && \widetilde{{\cal A}}_1(E)\equiv \widetilde{{\cal A}}_1(\widetilde{k})=
            \frac{1}{2}e^{-\widetilde{Q}_1a}
                   \left[
      \left(1+\frac{\widetilde{Q}_2}{\widetilde{Q}_1}\right) 
      e^{\widetilde{Q}_2a} \widetilde{{\cal A}}_3(\widetilde{k})+
     \left(1-\frac{\widetilde{Q}_2}{\widetilde{Q}_1} \right) 
      e^{-\widetilde{Q}_2a}  \widetilde{{\cal A}}_4(\widetilde{k})
     \right] , \hskip1.5cm  \\ 
     && \widetilde{ {\cal A}}_2(E)\equiv \widetilde{{\cal A}}_2(\widetilde{k})=
             \frac{1}{2}e^{\widetilde{Q}_1a}
                   \left[
      \left(1-\frac{\widetilde{Q}_2}{\widetilde{Q}_1}\right) 
      e^{\widetilde{Q}_2a} \widetilde{{\cal A}}_3(\widetilde{k})+
     \left(1+\frac{\widetilde{Q}_2}{\widetilde{Q}_1} \right) 
      e^{-\widetilde{Q}_2a}  \widetilde{{\cal A}}_4(\widetilde{k})
     \right] . \hskip1.5cm
\end{eqnarray}
     \label{tildeAfunc}
\end{mathletters}
If we define
\begin{equation}
      k:=\sqrt{\frac{2m}{\hbar ^2} \, E} \, ; \quad 
      Q_1:=\sqrt{\frac{2m}{\hbar ^2} \, (E+V_1)} \, ; \quad
      Q_2:=\sqrt{\frac{2m}{\hbar ^2} \, (E-V_2)} \, .
\end{equation}
then the functions ${\cal J}(E)$ of Eq.~(\ref{chi}) are given by
\begin{mathletters}
\begin{eqnarray}
      &&{\cal J}_1(E)\equiv {\cal J}_1(k)=
      \frac{1}{2}e^{-iQ_2a} 
     \left( \sin (Q_1a)+\frac{Q_1}{iQ_2}\cos (Q_1a) \right) , \hskip1.5cm \\
      &&{\cal J} _2 (E)\equiv {\cal J}_2(k)=
        \frac{1}{2}e^{iQ_2a} 
     \left( \sin (Q_1a)-\frac{Q_1}{iQ_2} \cos (Q_1a) \right) , \hskip1.5cm \\
      &&{\cal J}_3(E)\equiv {\cal J}_3(k)=
        \frac{1}{2}e^{-ikb}
                   \left[
            \left(1+ \frac{Q_2}{k} \right) 
       e^{iQ_2b}{\cal J}_1(k)+
       \left(1-\frac{Q_2}{k} \right) 
       e^{-iQ_2b}{\cal J}_2(k) 
        \right] , \hskip1.5cm \\
       &&{\cal J}_4(E)\equiv {\cal J}_4(k)=
         \frac{1}{2}e^{ikb}
                   \left[
            \left(1-\frac{Q_2}{k} \right) 
       e^{iQ_2b}{\cal J} _1(k)+
       \left(1+\frac{Q_2}{k}\right) 
        e^{-iQ_2b}{\cal J}_2(k) 
        \right] , \hskip1.5cm 
\end{eqnarray}
     \label{Jfunction}
\end{mathletters}
and the functions ${\cal A}^+(E)$ of Eq.~(\ref{theta+fun}) by
\begin{mathletters}
\begin{eqnarray}
     &&{\cal A}_3^+(E)\equiv {\cal A}_3^+(k)=
        \frac{1}{2}e^{-iQ_2b} 
     \left(1+\frac{k}{Q_2}\right) e^{ikb} , \hskip1.3cm \\
     &&{\cal A}_4^+(E)\equiv {\cal A}_4^+(k)=
         \frac{1}{2}e^{iQ_2b}
       \left(1-\frac{k}{Q_2} \right)
       e^{ikb} , \hskip1.3cm \\
     &&{\cal A}_1^+(E)\equiv {\cal A}_1^+(k)=
        \frac{1}{2}e^{-iQ_1a}
                   \left[
      \left(1+\frac{Q_2}{Q_1}\right) 
      e^{iQ_2a}{\cal A}_3^+(k)+
     \left(1-\frac{Q_2}{Q_1} \right) 
      e^{-iQ_2a} {\cal A}_4^+(k)
     \right] , \hskip1.3cm \\ 
     &&{\cal A}_2^+(E)\equiv {\cal A}_2^+(k)=
           \frac{1}{2}e^{iQ_1a}
                   \left[
      \left(1-\frac{Q_2}{Q_1}\right) 
      e^{iQ_2a}{\cal A}_3^+(k)+
     \left(1+\frac{Q_2}{Q_1} \right) 
      e^{-iQ_2a} {\cal A}_4^+(k)
     \right] . \hskip1.3cm
\end{eqnarray}
     \label{A+functions}
\end{mathletters}
The coefficients in Eq.~(\ref{thetafun-}) are given by
\begin{mathletters}
\begin{eqnarray}
    &&{\cal A}_3^-(E)\equiv {\cal A}_3^-(k)=
       \frac{1}{2}e^{-iQ_2b} 
     \left(1-\frac{k}{Q_2}\right) 
     e^{-ikb} , \hskip1.2cm \\
    &&{\cal A}_4^-(E)\equiv {\cal A}_4^-(k)=
      \frac{1}{2}e^{iQ_2b}
       \left(1+\frac{k}{Q_2} \right)
       e^{-ikb} , \hskip1.2cm \\
    &&{\cal A}_1^-(E)\equiv {\cal A}_1^-(k)=
       \frac{1}{2}e^{-iQ_1a}
                   \left[
      \left(1+\frac{Q_2}{Q_1}\right) 
      e^{iQ_2a}{\cal A}_3^-(k)+
     \left(1-\frac{Q_2}{Q_1} \right) 
      e^{-iQ_2a}{\cal A}_4^-(k)
     \right] , \hskip1.2cm \\ 
    &&{\cal A}_2^-(E)\equiv {\cal A}_2^-(k)=
         \frac{1}{2}e^{iQ_1a}
                   \left[
      \left(1-\frac{Q_2}{Q_1}\right) 
      e^{iQ_2a}{\cal A}_3^-(k)+
     \left(1+\frac{Q_2}{Q_1} \right) 
      e^{-iQ_2a}{\cal A}_4^-(k)
     \right]  . \hskip1.2cm
\end{eqnarray}
     \label{A-functions}
\end{mathletters}
The coefficients in Eq.~(\ref{tildesigma1}) are given by
\begin{mathletters}
\begin{eqnarray}
      && \widetilde{{\cal B}}_3(E)\equiv \widetilde{{\cal B}}_3(\widetilde{k})=
      \frac{1}{2}e^{-\widetilde{Q}_2b} 
     \left(1+\frac{\widetilde{k}}{\widetilde{Q}_2}\right)
     e^{\widetilde{k}b} , \\
     && \widetilde{{\cal B}}_4(E)\equiv \widetilde{{\cal B}}_4(\widetilde{k}) =
       \frac{1}{2}e^{\widetilde{Q}_2b}
       \left(1-\frac{\widetilde{k}}{\widetilde{Q}_2} \right)
       e^{\widetilde{k}b}  , \\
    && \widetilde{{\cal B}}_1(E)\equiv \widetilde{{\cal B}}_1(\widetilde{k})=
        \frac{1}{2}e^{-\widetilde{Q}_1a}
                   \left[
      \left(1+\frac{\widetilde{Q}_2}{\widetilde{Q}_1}\right) 
      e^{\widetilde{Q}_2a} \widetilde{{\cal B}}_3(\widetilde{k})+
     \left(1-\frac{\widetilde{Q}_2}{\widetilde{Q}_1} \right) 
      e^{-\widetilde{Q}_2a}  \widetilde{{\cal B}}_4(\widetilde{k})
     \right] , \\ 
     && \widetilde{ {\cal B}}_2(E)\equiv \widetilde{{\cal B}}_2(\widetilde{k})=
     \frac{1}{2}e^{\widetilde{Q}_1a}
                   \left[
     \left(1-\frac{\widetilde{Q}_2}{\widetilde{Q}_1} \right) 
     e^{\widetilde{Q}_2a} \widetilde{{\cal B}}_3(\widetilde{k})+
     \left(1+\frac{\widetilde{Q}_2}{\widetilde{Q}_1} \right) 
      e^{-\widetilde{Q}_2a} \widetilde{{\cal B}}_4(\widetilde{k})
     \right] .
\end{eqnarray}
     \label{tildeBfunctions}
\end{mathletters}
The functions ${\cal C}(E)$ of Eq.~(\ref{sigam2cos}) are given by
\begin{mathletters}
\begin{eqnarray}
      &&{\cal C}_1(E)\equiv {\cal C}_1(k)= 
       \frac{1}{2}e^{-iQ_2a}
             \left( \cos (Q_1a)-\frac{Q_1}{iQ_2} \sin (Q_1a) \right), \\
     && {\cal C}_2(E)\equiv {\cal C}_2(k)=
       \frac{1}{2}e^{iQ_2a}
        \left( \cos (Q_1a)+\frac{Q_1}{iQ_2} \sin (Q_1a) \right), \\
     && {\cal C}_3(E)\equiv {\cal C}_3(k)=\frac{1}{2}e^{-ikb}
                   \left[
      \left(1+\frac{Q_2}{k}\right) e^{iQ_2b}{\cal C}_1(k)+
     \left(1-\frac{Q_2}{k} \right) e^{-iQ_2b}{\cal C}_2(k)
     \right] , \\ 
     &&{\cal C}_4(E)\equiv {\cal C}_4(k)=\frac{1}{2}e^{ikb}
                   \left[
     \left(1-\frac{Q_2}{k} \right) e^{iQb}{\cal C}_1(k)+
     \left(1+\frac{Q_2}{k} \right) 
      e^{-iQb}{\cal C}_2(k)
     \right]  . 
\end{eqnarray}
     \label{Cfunctions}
\end{mathletters}

\def\thesection{\Alph{section}}
\section{Proofs of Propositions 1 and 2}
\setcounter{equation}{0}
\label{sec:appendixB}

{\it Proof of Proposition~1.} 

\vskip0.3cm

(i) It is very easy to show that the quantities (\ref{nmnorms}) fulfill the 
conditions to be a norm,
\begin{mathletters}
\begin{eqnarray}
      &&\| \varphi _{\rm c}+\psi _{\rm c} \| _{n,m} \leq 
       \| \varphi _{\rm c} \| _{n,m} + 
       \| \psi _{\rm c} \| _{n,m} \, , \\
      && \| \alpha \varphi _{\rm c} \| _{n,m}=
          |\alpha |\, \| \varphi _{\rm c}\| _{n,m} \, , \\
      && \| \varphi _{\rm c} \| _{n,m} \geq 0 \, , \\
      && {\rm If }\  \| \varphi _{\rm c} \| _{n,m} =0, \ {\rm then} \ 
          \varphi _{\rm c} =0 \, .
        \label{homiensi}
\end{eqnarray}
\end{mathletters}
The only condition that is somewhat difficult to prove is (\ref{homiensi}): if 
$\| \varphi _{\rm c}\| _{n,m}=0$, then 
\begin{equation}
       (1+r)^n(h+1)^m\varphi _{\rm c}(r)=0 \, ,
\end{equation}
which yields
\begin{equation}
      (h+1)^m\varphi _{\rm c}(r)=0 \, .  
      \label{homiodhiis}
\end{equation}
If $m=0$, then Eq.~(\ref{homiodhiis}) implies $\varphi _{\rm c}(r)=0$.  If 
$m=1$, then Eq.~(\ref{homiodhiis}) implies that $-1$ is an eigenvalue of 
$H_{\rm c}$ whose corresponding eigenvector is $\varphi _{\rm c}$.  Since 
there is no discrete eigenvalue in the continuous part of the spectrum, 
$\varphi _{\rm c}$ must be the zero vector.  The proof for $m>1$ is similar.

\vskip0.3cm

(ii) In order to see that $H_{\rm c}$ is 
$\tau _{\mathbf \Phi _{\rm c}}$-continuous, we just have to realize that
\begin{eqnarray}
      \| H_{\rm c}\varphi _{\rm c}\| _{n,m}&=&
     \| (H_{\rm c}+I)\varphi _{\rm c}-\varphi _{\rm c}\| _{n,m} \nonumber \\
       &\leq & \| (H_{\rm c}+I)\varphi _{\rm c}\| _{n,m}+ 
        \| \varphi _{\rm c}\| _{n,m} \nonumber \\
       &=&\| \varphi _{\rm c}\| _{n,m+1}+\| \varphi _{\rm c}\| _{n,m} \, .
       \label{tauphiscont}
\end{eqnarray}
The stability of $\mathbf \Phi _{\rm c}$ under the action of 
$H_{\rm c}$ also follows from Eq.~(\ref{tauphiscont}).

\vskip0.3cm

(iii) From the definition (\ref{definitionket}), it is pretty easy 
to see that $|E\rangle$ is an antilinear functional.  In order to show that 
$|E\rangle$ is continuous, we define
\begin{equation}
      {\cal M}(E):= \sup _{r\in [0,\infty )} \left| \phi (r;E) \right| \, .
\end{equation}
Since
\begin{eqnarray}
      |\langle \varphi _{\rm c}|E\rangle | &=&\left| \int_0^{\infty}dr \, 
         \overline{\varphi _{\rm c}(r)}\phi(r;E)\right| \nonumber \\
     &\leq & {\cal M}(E) \int _0^{\infty}dr \, |\varphi _{\rm c}(r)|
           \nonumber \\
      &=& {\cal M}(E) \int_0^{\infty}dr \,
      \frac{1}{1+r} (1+r) |\varphi _{\rm c}(r)| \nonumber \\
      &\leq & {\cal M}(E) \left( \int_0^{\infty}dr \, 
      \frac{1}{(1+r)^2} \right) ^{1/2} 
      \left( \int_0^{\infty}dr \, 
      \left| (1+r) \varphi _{\rm c}(r) \right| ^2 \right) ^{1/2} \nonumber \\
      &=&{\cal M}(E) \| \varphi _{\rm c} \| _{1,0} \, ,
\end{eqnarray}
the functional $|E\rangle$ is continuous when $\mathbf \Phi _{\rm c}$ is 
endowed with the $\tau _{\mathbf \Phi _{\rm c}}$ topology.

\vskip0.3cm

(iv) In order to prove that $|E\rangle$ is a generalized eigenvector of 
$H_{\rm c}$, we make use of the conditions (\ref{mainisexi}) and 
(\ref{condcodogis}) satisfied by the elements of $\mathbf \Phi _{\rm c}$,
\begin{eqnarray}
       \langle \varphi _{\rm c}|H_{\rm c}^{\times}|E\rangle &=& 
      \langle H_{\rm c}^{\dagger}\varphi _{\rm c}|E\rangle
       \nonumber \\
       &=& \int_0^{\infty}dr \, 
       \left( -\frac{\hbar ^2}{2m}\frac{d^2}{dr^2}+V(r) \right)
       \overline{\varphi _{\rm c}(r)} \phi (r;E) \nonumber \\
       &=&-\frac{\hbar ^2}{2m}
       \left[ \frac{d\overline{\varphi _{\rm c}(r)}}{dr} \phi (r;E)
       \right] _0^{\infty} 
       +\frac{\hbar ^2}{2m}
       \left[ \overline{\varphi _{\rm c}(r)} \frac{d\phi(r;E)}{dr} 
       \right] _0^{\infty} \nonumber \\ 
       &&+ \int_0^{\infty}dr \, \overline{\varphi _{\rm c}(r)}
       \left( -\frac{\hbar ^2}{2m}\frac{d^2}{dr^2}+V(r) \right)\phi (r;E)
        \nonumber \\
       &=&E\langle \varphi _{\rm c}|E\rangle \, .
\end{eqnarray}

\vskip0.5cm

{\it Proof of Proposition~2}. 

Let $\varphi$ and $\psi$ be in $\mathbf \Phi$.  Since $U$ of 
Eq.~(\ref{dirintdec}) is unitary,
\begin{equation}
       (\varphi ,\psi )=(U\varphi ,U\psi )=
       (\widehat{\varphi}_{\rm b} ,\widehat{\psi}_{\rm b})_{{\cal H}_{\rm b}}+
       (\widehat{\varphi}_{\rm c} ,\widehat{\psi}_{\rm c})_{{\cal H}_{\rm c}}
       \, .
       \label{Usiuni}
\end{equation}
The scalar product of $\widehat{\varphi}_{\rm b}$ and $\widehat{\psi}_{\rm b}$
can be written as
\begin{equation}
      (\widehat{\varphi}_{\rm b} ,\widehat{\psi}_{\rm b})_{{\cal H}_{\rm b}}=
      \sum_{n=1}^N(\varphi|E_n)(E_n|\psi ) \, .
      \label{discreslsl}
\end{equation}
The wave functions $\widehat{\varphi}_{\rm c}$ and $\widehat{\psi}_{\rm c}$ 
are in particular elements of $L^2([0,\infty ),dE)$.  Hence their scalar 
product is well-defined,
\begin{equation}
      (\widehat{\varphi}_{\rm c} ,\widehat{\psi}_{\rm c})_{{\cal H}_{\rm c}}=
      \int_0^{\infty}dE \, 
      \overline{ \widehat{\varphi}_{\rm c}(E)}\, \widehat{\psi}_{\rm c}(E) \, .
      \label{sphatvhaps}
\end{equation}
Since $\varphi _{\rm c}$ and $\psi _{\rm c}$ belong to 
$\mathbf \Phi _{\rm c}$, the action of each eigenket $|E\rangle$ on them is 
well-defined,
\begin{mathletters}
\begin{eqnarray}
      \langle \varphi _{\rm c}|E\rangle =
            \overline{ \widehat{\varphi}_{\rm c}(E)} \, ,\\
      \langle E|\psi _{\rm c}\rangle =\widehat{\psi}_{\rm c}(E) \, .
\end{eqnarray}
        \label{actionofEpsi}
\end{mathletters}
By plugging Eq.~(\ref{actionofEpsi}) into Eq.~(\ref{sphatvhaps}) and
Eqs.~(\ref{sphatvhaps}) and (\ref{discreslsl}) into Eq.~(\ref{Usiuni}), 
we get to Eq.~(\ref{GMT1}).  The proof of (\ref{GMT2}) follows the same 
pattern.

\end{document}